\documentclass[8.5pt,twoside,twocolumn]{article}
\usepackage[super,sort&compress,comma]{natbib} 
\usepackage{url}
\usepackage{mhchem}
\usepackage{times,mathptmx}
\usepackage{sectsty}
\usepackage{balance} 
\usepackage{graphicx} 
\usepackage{lastpage}
\usepackage[format=plain,justification=raggedright,singlelinecheck=false,font=small,labelfont=bf,labelsep=space]{caption} 
\usepackage{fancyhdr}
\usepackage{multirow}
\usepackage{float}

\begin{document}
\thispagestyle{plain}
\fancypagestyle{plain}{
\renewcommand{\headrulewidth}{1pt}}
\renewcommand{\thefootnote}{\fnsymbol{footnote}}
\renewcommand\footnoterule{\vspace*{1pt}%
\hrule width 3.4in height 0.4pt \vspace*{5pt}} 
\setcounter{secnumdepth}{5}
\makeatletter 
\def\subsubsection{\@startsection{subsubsection}{3}{10pt}{-1.25ex plus -1ex minus -.1ex}{0ex plus 0ex}{\normalsize\bf}} 
\def\paragraph{\@startsection{paragraph}{4}{10pt}{-1.25ex plus -1ex minus -.1ex}{0ex plus 0ex}{\normalsize\textit}} 
\renewcommand\@biblabel[1]{#1}            
\renewcommand\@makefntext[1]%
{\noindent\makebox[0pt][r]{\@thefnmark\,}#1}
\makeatother 
\renewcommand{\figurename}{\small{Fig.}~}
\sectionfont{\large}
\subsectionfont{\normalsize} 

\fancyfoot{}
\fancyhead{}
\renewcommand{\headrulewidth}{1pt} 
\renewcommand{\footrulewidth}{1pt}
\setlength{\arrayrulewidth}{1pt}
\setlength{\columnsep}{6.5mm}
\setlength\bibsep{1pt}

\twocolumn[
  \begin{@twocolumnfalse}
\noindent\LARGE{\textbf{Efficiency enhancement of black dye-sensitized solar cell by newly synthesized D-$\pi$-A coadsorbents: A theoretical study}}
\vspace{0.6cm}

\noindent\large{\textbf{Yavar T. Azar and Mahmoud Payami$^{\ast}$}}\vspace{0.5cm}

\vspace{0.6cm}

\noindent \normalsize{In this work, using the DFT and TDDFT, we have theoretically studied the electronic and optical properties of the two recently synthesized coadsorbents Y1 and Y2, which were aimed to enhance the efficiency of the black dye-sensitized solar cells. To determine the solvatochromic shifts, both the implicit and mixed implicit-explicit models have been used. The connection between the solvatochromic shifts and the changes of dipole moments in the excitation process is discussed. The difference in excitation charge transfer is utilized to explain the experimentally observed difference in $J_{sc}$ for Y1 and Y2. Investigating the interactions of I$_2$ molecules in the electrolyte solution with the coadsorbents showed that with Y1 the recombination loss was weakened through decreasing the I$_2$ concentration near the TiO$_2$ surface, whereas with Y2 it was increased. As a result, the higher values of both $J_{sc}$ and $V_{oc}$ with Y1 coadsorbent explains its experimentally observed higher efficiency. The present study sheds light on how to design and engineer newer coadsorbents or organic dyes for higher efficiencies.}
\vspace{0.5cm}
 \end{@twocolumnfalse}
  ]

\section{Introduction}\label{sec1}

\footnotetext{\textit{Theoretical and Computational Physics Group, School of Physics and Accelerators, AEOI, P.~O.~Box~14395-836, Tehran, Iran; E-mail: mpayami@aeoi.org.ir}}

The advent of dye-sensitized solar cells (DSSCs), so-called Gratzel cells,\cite{Oregan1991} in the early nineties had drastic effects on the photovoltaic technology. Because of low-cost materials and easy fabrication, DSSC technology is one of the promising alternatives for commercial solar energy conversion. In a typical DSSC, the light-excited dye molecules, which are adsorbed on TiO$_2$ nano-particles, inject the excited electrons into the lower unoccupied electron states in the conduction band (CB) of the TiO$_2$ semiconductor. The injected electrons move through the load to the counter-electrode and regenerate I$^-$ by the reduction of I$^-_3$, while the ionized dye is reduced\cite{hagfeldt2010} by I$^-$.

The efficiency of a DSSC can be improved mainly in two ways: (i) designing and synthesizing of new materials, (ii) reducing unfavored loss mechanisms.\cite{oregan2009,Pal02,oregan2008,Maggio2013} The most important loss mechanism, which has significant impact on the performance of the DSSC, is the charge recombination in which the injected electrons are back-transferred from the semiconductor to the oxidized dye or are captured by the molecules in the electrolyte solution. The loss mechanism which involves the semiconductor and the dye molecule, can be reduced by designing new dye materials. To overcome the recombinations at the electrolyte-TiO$_2$ interface, the surface passivation by different organic molecules or utilizing some additives in the electrolyte solution had already been proposed.\cite{Lee2012,Zhang2012} Another loss mechanism, which is the light absorption by the triiodide in the electrolyte (competitive light absorption), leads to a significant reduction in the incident photon-to-current conversion efficiency (IPCE).\cite{Boschloo2009} For example, a 1 M concentration of triiodide in electrolyte gives rise to 13$\%$  photocurrent reduction in a DSSC with standard N719 dye.\cite{Wang2003}

\begin{figure}
\includegraphics[width=\columnwidth]{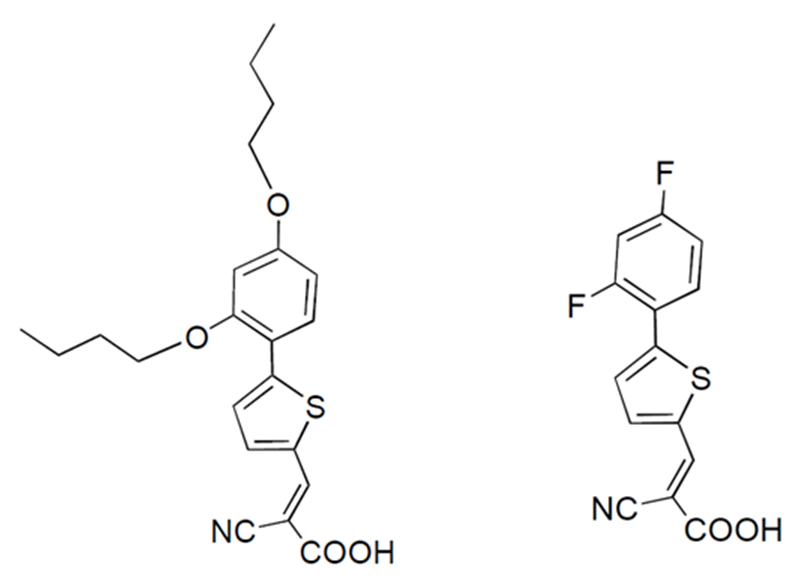}
\caption{The sketch map structures of Y1 (left) and Y2 (right).}
\label{fig1}
\end{figure}

Recently, two organic coadsorbents, Y1 and Y2, had been designed (see Fig.~\ref{fig1}) and synthesized\cite{Han2012} expecting that when coadsorbed with black dye (also known as N749) on TiO$_2$ surface, (i) they prevent the aggregation of the dye molecules on the TiO$_2$ surface, (ii) operate as additional organic sensitizers for light harvesting in blue and UV regions and thereby preventing the light absorption by triiodide, and (iii) blocking the electron captures by the molecules in the electrolyte solution. 
In these molecules, the cyanoacetic groups act as acceptors, the thiophenes act as  $\pi$-spacers, and the remaining parts act as donor units.  
Utilizing these coadsorbents with black dye had shown that Y1 increases the efficiency by 0.58$\%$ while Y2 decreases it by 0.42$\%$ (See Table 1 in Ref.~\citenum{Han2012}). In this way, the coadsorption of Y1 with the black dye leads to a new record of efficiency 11.4$\%$ in the family of black dye sensitized solar cells. 

In this work, using the density functional theory\cite{HK64} (DFT) and the time-dependent density functional theory\cite{RG84} (TDDFT) within the B3LYP\cite{B3LYP93-1,B3LYP93-2} and CAM-B3LYP\cite{Yanai2004} approximations for the exchange-correlations (XC), we have theoretically studied the electronic structure, the optical properties, and the adsorption geometry of these coadsorbents both in vacuum and in solvents. The solvent effects on the absorption spectra (solvatochromic effects), are calculated for the cases when Y1 and Y2 are embedded in the  acetonitrile (AN) or in the ethanol (EtOH) media. Acetonitrile is chosen because it was the most utilized\cite{hagfeldt2010} solvent in DSSCs and EtOH is the only available reported solvent that was experimentally used\cite{Han2012} for Y1 and Y2. 

When Y1 and Y2 are deposited on the semiconductor surface, it is more likely that the hydrogen atom of the carboxylic acid dissociates (deprotonation) and the bond forms between the carboxylic oxygen atoms and the TiO$_2$ surface. Deprotonation is likely to happen also when Y1 and Y2 are embedded in a solvent. We, therefore, have considered deprotonation effects in the calculation of the electronic structure and the optical properties. As will be discussed in the following, the calculation results show that deprotonations of Y1 and Y2 by ethanol lead to significant blue shifts in UV/Vis spectra. 

Finally, to understand the origins of the efficiency-enhancement by Y1 and the efficiency-drop by Y2, we checked the possibility of loss via back-transfer mechanism by analyzing the charge distributions of the HOMO orbitals for the neutral Y1 and Y2 molecules\cite{Troisi2012} and we found out that it could not be the main reason for efficiency difference. We were therefore led to the investigation of loss mechanism of electron-capture by I$_2$ molecules in the electrolyte solution.
 
In the absence of the coadsorbents, the I$_2$ molecules have a finite probability to reach the semiconductor surface to capture the electrons. Our results show that in the Y2 case, formation of halogen bonds between the I$_2$ molecules and CN group increases the recombination rate; while in the Y1 case, the butyloxyl groups keep the I$_2$ molecules away from the surface, which in turn, decreases the recombination rate through the electron capture. That is, keeping the other parameters more or less the same, the loss mechanism in the Y2 case is strengthened (leading to efficiency drop) while it is weakened in the Y1 case leading to the efficiency enhancement. 

In section~\ref{sec2}, we present the computational details. Section~\ref{sec3} is devoted to the calculated results and discussion, and finally, in section~\ref{sec4} we conclude this work. 
\section{Computational details}\label{sec2}
Equilibrium geometries of neutral Y1 and Y2 as well as their corresponding deprotonated anions were determined using GAMESS package,\cite{Schmidt1993} within B3LYP approximation for the XC functional and 6-31+G(d) basis set for both gas phase and in solvent. 

Due to its proper treatment of the long-range electrostatic effects of the solvents (polarization effects), we have employed the polarized continuum model\cite{Cramer1999,Tomasi2005} (PCM) in which the solvent is assumed to be a structureless dielectric medium and the solute is confined in a cavity comprised of overlapping spheres centered on atoms. Among the available developments of continuum model, we have used the most popular and fast one, the conductor-like PCM\cite{Klamt93} (C-PCM). In this model, the surface charge density, which is obtained by assuming the surrounding medium as a conductor ($\epsilon$=$\infty$), is renormalized by a scaling function $f(\epsilon)$, 
\begin{equation}\label{eq1}
f(\epsilon)=\frac{\epsilon-1}{\epsilon+k}
\end{equation}
to give accurate charge density for the real medium with finite dielectric constant. The values 25 and 36 are assumed for the dimensionless $\epsilon$ of EtOH and AN, respectively. In the GAMESS code, the default value of $k=0$ is assumed in the denominator of equation~(\ref{eq1}). 

The OH group in the molecular structure of ethanol can form hydrogen bonding with the carboxylic moiety and other electronegative atoms of coadsorbent. In this case, the C-PCM in which the short-range interactions between solute and solvent are neglected, is not adequate and may lead to inaccurate results. We have therefore used a generalized continuum model, the so-called solvated supermolecule method (implicit-explicit method), in which the solute molecule is augmented by a few solvent molecules (usually those belonging to the first solvation shells), and the resulting supermolecule is treated by C-PCM.\cite{Cappelli2000,Tomasi2005,Marenich2011} We have applied this method by adding six ethanol molecules to the carboxylic group and high electronegative sites of Y1 and Y2 (see Fig.~\ref{fig3}). 

The excitation energies as well as the oscillator strengths were calculated from solving the Casida equations\cite{Casida1995,Dreuw2005,Casida2009} as implemented in GAMESS code. 
It has been shown\cite{Dreuw2004} that the TDDFT calculations with the B3LYP hybrid functional gives relatively accurate results only when the excitations does not have a significant charge-transfer (CT) character, whereas for CT-character excitations, a newer hybrid functional, CAM-B3LYP, is more appropriate.\cite{Yanai2004,adamo2013} In these calculations, the B3LYP and range-separated CAM-B3LYP approximations were used.
 
To have a meaningful comparison with experiment, the resulting oscillator strengths are convoluted by Gaussian functions with an appropriate full width at half maximum, $\Delta_{1/2}$, to obtain the extinction coefficient as
\begin{equation}\label{eq2}
\epsilon(\omega)=2.174 \times10^{8}\sum_I{\frac{f_I}{\Delta_{1/2}}exp[\;2.773\frac{(\omega_I^2-\omega^2)^2}{\Delta_{1/2}^2}\;]}
\end{equation}
where, $f_I$ and $\omega_I$ are the $I$th oscillator strength and excitation frequency, respectively. 

For the excited-state calculations of Y1 and Y2 in solvent, we used non-equilibrium C-PCM/TDDFT which is based on the fact that the response of the solvent electrons to the "instantaneous" change of the solute charge distribution (due to the excitation) is very fast compared to that of the ions.\cite{Cossi2001,Marenich2011}
In the calculation of vertical excitation energies, only the electronic response was considered and the motion of the solvent ions were assumed to be frozen.\cite{adamo2013}

To determine the deposition geometry of Y1 and Y2 on the surface of TiO$_2$ nanoparticles, we have used both the periodic-slab and cluster methods. In the periodic-slab method, we have constructed an anatase 4-(TiO$_2$)-layer slab with (101) surface using a $4\times 1$ supercell along [010] and [10$\bar{1}$] directions. These calculations are based on the DFT and the self-consistent solution of the Kohn-Sham (KS) equations\cite{KS65} using the Quantum ESPRESSO code package\cite{QE-2009} within the PBE generalized gradient approximation\cite{PBE96} for the XC energy functional. For the atoms Ti, O, C, N, S, F, and H  we have used the pseudopotentials Ti.pbe-sp-van\_ak.UPF, O.pbe-van\_bm.UPF, C.pbe-van\_bm.UPF, N.pbe-van\_ak.UPF, S.pbe-van\_bm.UPF, F.pbe-n-van.UPF, and H.pbe-van\_ak.UPF from http://www.quantum-espresso.org. The kinetic-energy cutoff for the plane-wave basis set were 28 and 220 Ryd for the wave functions and charge density, respectively. For the Brillouin-zone integrations, a $2\times2\times1$ grid was used. In the cluster calculations, because of yielding results in good agreement with experiment, we used an anatase (TiO$_2)_{38}$ cluster to model the nanoparticles.\cite{Persson2000,Pastore2012chem,deangelis2008}

Geometry optimization of  I$_2$-Y1 and I$_2$-Y2 adducts were performed using NWChem code package\cite{Valiev2010} with the  6-311G** basis set within B3LYP approximation. The interaction energies between I$_2$ and the coadsorbents Y1/Y2 were calculated using
\begin{equation} \label{eq3}
E_{\rm int}=E_{\;\rm I_2\cdots\;Y1/Y2}-(E_{\;\rm I_2}+E_{\;\rm Y1/Y2})-\Delta E_{\;\rm CP} 
\end{equation}
in which $E_{\;\rm I_2\cdots\;Y1/Y2}$ is the total energy of the I$_2$-Y1 or I$_2$-Y2 adducts, $E_{\;\rm I_2}$ and $E_{\;\rm Y1/Y2}$ are the total energies of the isolated components, and $\Delta E_{\;\rm CP}$ is the compensation correction for the superposition error in the basis set.\cite{boys1970}
\section{Results and discussion}\label{sec3}
\subsection{Equilibrium properties of Y1 and Y2}
\begin{figure}[h]
\includegraphics[width=\columnwidth]{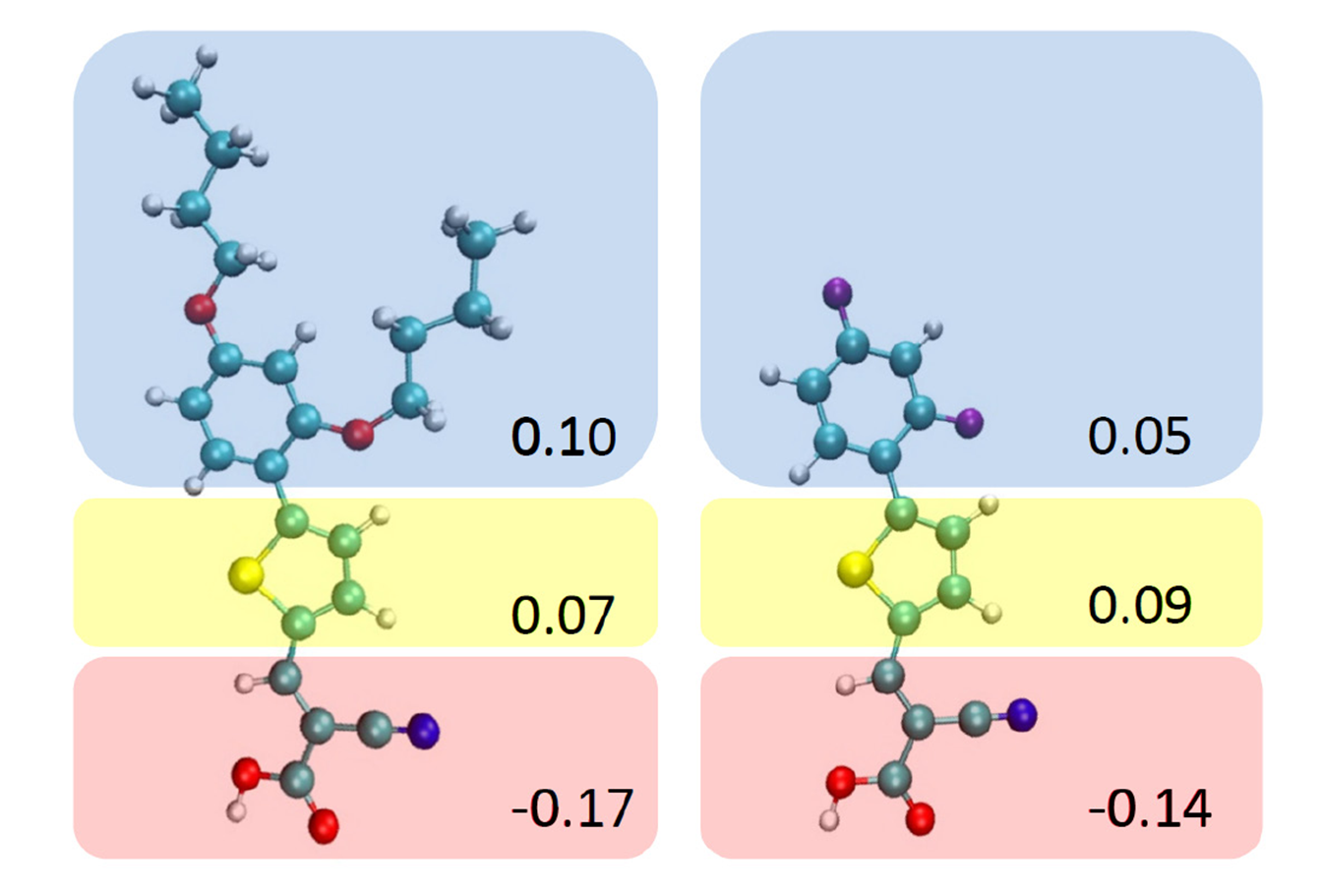}
\caption{Optimized geometrical structures of Y1 (left) and Y2 (right) in vacuum within B3LYP. The red, dark blue, yellow, and violet balls correspond to O, N, S, and F atoms, respectively. Small balls represent H atoms, and all others are C atoms. The L\"owdin charges on the donor (top), spacer (middle), and acceptor (bottom) units are specified.}
\label{fig2}
\end{figure}

Full optimization of geometrical structures has been carried out for Y1 and Y2 molecules in both vacuum and solvent using GAMESS within B3LYP/6-31+G(d) basis set, and the result for vacuum is shown in Fig.~\ref{fig2}. 

The donor, spacer, and acceptor units of the molecules are specified from top to bottom of the figure, respectively. The dihedral angles, in degrees, and bond lengths, in $\AA$, for acceptor-spacer (A-S) and spacer-donor (S-D) in gas phase (vacuum) and in solvents are presented in Table~\ref{tab1}.

\begin{table}[h]
  \caption{Bond lengths in $\AA$ and dihedral angles in degrees for A-S and S-D groups. "Dep" stand for deprotonated in ethanol, and "Sup" stand for supermolecule. }
\small\center
\begin{tabular}{@{}llcccc} \hline \hline 
            & &\;\;$d_{\rm A-S}$\;\;&\;\;$\theta_{\rm A-S}$\;\; &\;\;$d_{\rm S-D}$\;\;&\;\; $\theta_{\rm S-D}$\;\;  \\ \hline
\multirow{5}{*}{Y1}\;\;\;\;\; & Gas\;\;\;\;\; & 1.42 & -0.9 & 1.46 & -29.9  \\
                    & EtOH& 1.42 & -1.1 & 1.46 & -20.5  \\ 
                    & Dep & 1.44 &  0.5 & 1.46 & -37.1      \\
    	              & AN& 1.42 & -1.2  & 1.46 & -21.5 \\ 
                    & Sup & 1.42 & -1.4 & 1.46 & -12.7      \\ \hline
\multirow{5}{*}{Y2} & Gas & 1.43 & 0.1  & 1.46 & -26.9    \\
                    & EtOH& 1.42 & 0.6  & 1.46 & -29.3   \\ 
                    & Dep & 1.44 &  0.8 & 1.46 & -29.4      \\
    	            & AN  & 1.42 & 0.7  & 1.46 & -29.2   \\ 
                     & Sup & 1.42 & 2.8  & 1.46 & -29.2      \\ \hline
\end{tabular}
\label{tab1}
\end{table}

\begin{figure}[h]
\includegraphics[width=\columnwidth]{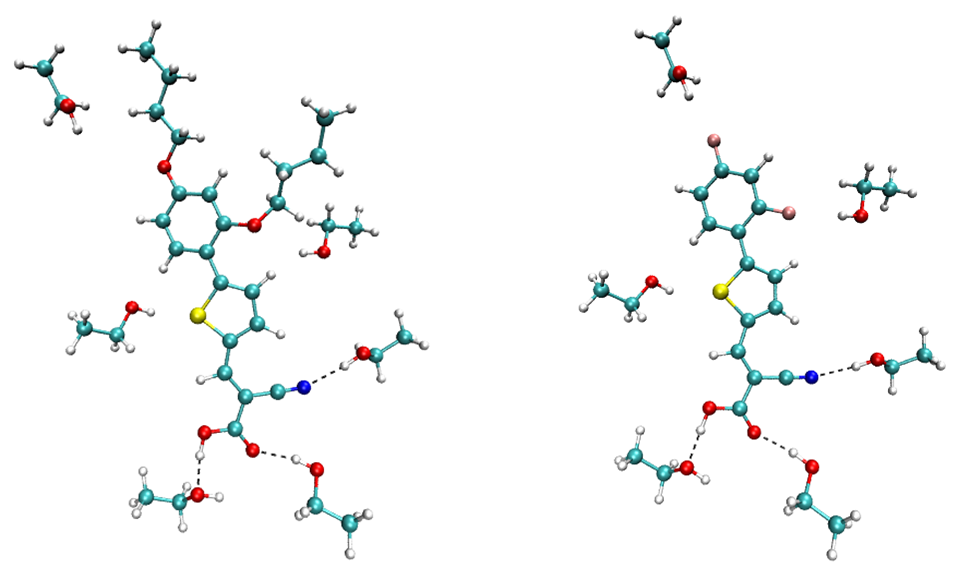}
\caption{Optimized geometrical structures of Y1 (left) and Y2 (right) supermolecules. Dashed lines correspond to intermolecular bond lengths which are less than 2.5~$\AA$.}
\label{fig3}
\end{figure}

Inspecting the distances listed in Table~\ref{tab1}, and knowing the fact that Y2 was resulted from the substitution of butyloxyl with fluorine atoms in Y1, we conclude that: (i) these substitutions do not affect the A-S or S-D bond lengths, (ii) the solvents do not affect these parameters compared to gas phase, and (iii) deprotonated molecules are the only cases that show  small increase in the A-S distances. The small dihedral angles, $\theta_{\rm A-S}$, show that the thiophene unit and cyanoacetic group are coplanar which, in turn, leads to a strong conjugation across the combined A-S group. On the other hand, the results show that the dihedral angle of deprotonated Y1 is larger than that of the neutral one; whereas, in deprotonated Y2, the dihedral angle does not change compared to its neutral counterpart. Inspection of the other geometric parameters show that deprotonation of both Y1 and Y2 increases the bond length between carboxylic acid and its adjacent carbon atom from 1.48 to 1.54~$\AA$. Finally, using the supermolecular method (see Fig.~\ref{fig3}), the OH bond length of carboxylic acid increases from 0.97 to 1.01~$\AA$ due to the formation of hydrogen bonding with one of explicit ethanol molecules in the solvent, while the five other explicit ethanol molecules have no significant effects on the bond lengths. The bond lengths which are less than 2.5~$\AA$, are shown by dashed lines. 

\subsection{Optical properties and $J_{\rm sc}$}
To understand the charge transfer mechanism of the coadsorbents in different media, we calculated the ground- and excited-state L\"owdin charges on the acceptor, spacer, and donor units. Moreover, to explain the solvatochromic shifts, the molecular dipole moments were calculated both in vacuum and in solvent. The details are discussed separately in the following subsections.
\subsubsection{Ground state}
\begin{figure}[h]
\includegraphics[width=\columnwidth]{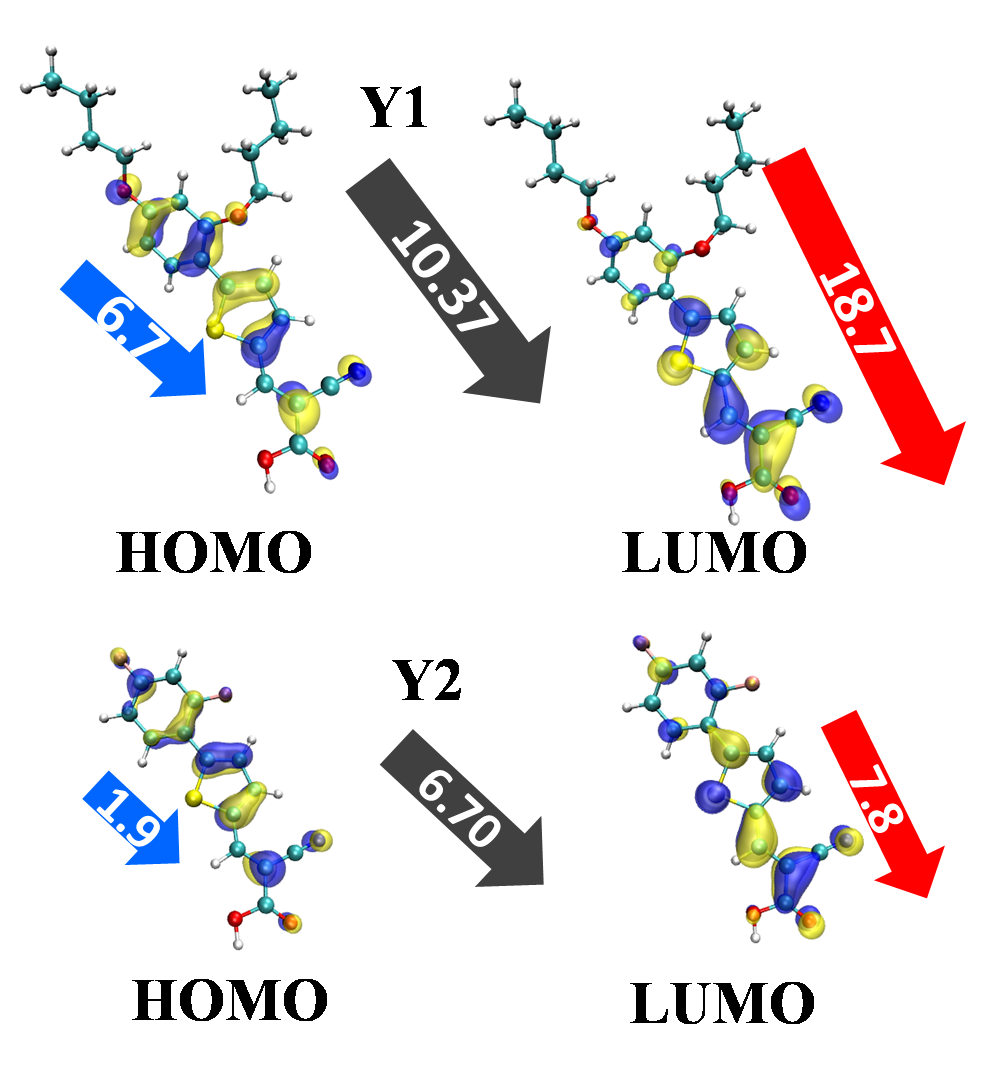}
\caption{HOMO and LUMO isodensity plots for optimized geometries of Y1 and Y2 in vacuum within B3LYP approximation. The middle arrows (black) show the magnitude and orientation of total dipole moments of the molecules. The left arrows (blue) and right arrows (red) represent the HOMO and LUMO dipole moments, respectively.}
\label{fig4}
\end{figure}

$\;$  
L\"owdin charge analysis for the equilibrium geometries (Table~\ref{tab2}) shows that for Y1, some electronic charges of the donor and acceptor units have been resided on the acceptor unit in all three gas, solvent, and deprotonated cases. 
\begin{table*}
\small \center
 \caption{Lowdin charges, in atomic units, on different groups in Y1 and Y2 molecules. The dipole moments in Debye units are listed in the last column. }
  {\begin{tabular}{@{}lllccccccccc} \hline \hline 
                   &  &  &\multicolumn{4}{c}{Ground state}&& \multicolumn{4}{c}{Excited state}  \\ \cline{4-7}  \cline{9-12}   
 & & &\multicolumn{3}{c}{L\"owdin charge (e) }&\;\;\;\; $\mu$ (Debye) &\;\;\;\;\;& \multicolumn{3}{c}{L\"owdin charge (e) }&\;\;\;$\mu$ (Debye) \\ \cline{4-6} \cline{9-11}
    &    & \;\;\;\;\;& Acceptor    & Spacer    &   Donor   &              && Acceptor   & Spacer       & Donor     &         \\ \hline
\multirow{6}{*}{Y1}\;\;\;\;\;&\multirow{2}{*}{Gas} & B3LYP& -0.18& 0.07& 0.11& 10.37 &&-0.36& -0.03& 0.39& 19.04\\
                   &                 &CAM-B3LYP\;\;\;\;\; & -0.15& 0.07& 0.09& 9.26 &&-0.32& 0.07& 0.25 & 15.28  \\ \cline{2-12}
	           &\multirow{2}{*}{Sol}& B3LYP& -0.24& 0.08& 0.16& 14.23 &&-0.49& -0.02& 0.51& 26.01 \\
                   &                 & CAM-B3LYP & -0.22& 0.08& 0.14& 13.26&&-0.46& 0.05& 0.41 & 23.01  \\ \cline{2-12}
                   &\multirow{2}{*}{Dep}& B3LYP& -1.06& 0.00  & 0.06  & 43.63 && -1.27 & -0.06  & 0.34  & 53.12  \\
                   &                 & CAM-B3LYP & -1.05  & 0.00  & 0.05  & 43.47  && -1.22  & 0.02  & 0.20  & 49.75  \\ \hline
\multirow{6}{*}{Y2}&\multirow{2}{*}{Gas} &B3LYP& -0.14& 0.09& 0.05&  6.70 &&-0.27& 0.09& 0.18 & 10.22  \\
                   &                 & CAM-B3LYP & -0.12& 0.07& 0.05&  6.35  &&-0.25& 0.12& 0.13 &  9.35  \\ \cline{2-12}
                   &\multirow{2}{*}{Sol} &B3LYP& -0.22& 0.21& 0.01&  9.69  &&-0.44& 0.23& 0.21 & 16.38  \\
                   &                 & CAM-B3LYP & -0.20& 0.19& 0.01&  9.36  &&-0.41& 0.29& 0.12 & 14.35  \\ \cline{2-12}
                   &\multirow{2}{*}{Dep} &B3LYP& -1.09  & 0.12  & -0.03  & 30.64   && -1.20  & 0.15  & 0.05   & 33.82  \\
                   &                 & CAM-B3LYP & -1.08  & 0.12  & -0.04  & 30.56   && -1.23  & 0.20   & 0.03   & 34.52  \\ \hline
      \end{tabular}}
   \label{tab2}
 \end{table*}
This charge separation leads to the formation of molecular dipole moment directed from donor to acceptor unit (see black arrows in Fig.~\ref{fig4}). However, for Y2, the electronegative fluorine atoms tend to prevent the charge separation in the donor unit and the negative charges on the acceptor are mostly supplied from the spacer unit. This explains why the molecular dipole moment of Y2 is smaller than that of Y1.  
The larger values of the dipole moment for deprotonated cases stem from the fact that the excess negative charges are almost completely accumulated on the acceptor groups leading to additional strong dipole moments. The values in Table~\ref{tab2} imply that both B3LYP and CAM-B3LYP predict more or less the same results for the L\"owdin charge distribution in the ground state. 

Since, in the C-PCM, the dependence of the solvent reaction field to the dielectric constant comes from the scaling function [Eq.~(\ref{eq1})], and the value of this function is approximately the same for AN and EtOH, only the results for EtOH is presented in Table~\ref{tab2}. 

The isodensity plots for HOMO and LUMO, as shown in Fig.~\ref{fig4}, indicate that for both Y1 and Y2, the HOMOs are significantly localized at the donor and spacer, while the LUMOs are confined to the spacer and acceptor units.

The HOMO and LUMO energy levels and the gap in between for the optimized structures of Y1 and Y2 are listed in Table~\ref{tab3}. 
Comparing with the gas-phase results show that, for Y1 in ethanol the HOMO level is slightly shifted upward while the LUMO energy is shifted downwards with the net effect that the HOMO-LUMO gap in ethanol is less than that in vacuum by 0.17~eV. Similar analysis for Y2 revealed that both the HOMO and LUMO levels had shifted upwards but since the HOMO level-shift was greater than the LUMO level-shift, the net effect was a narrower gap for Y2 by 0.07~eV. These results can not be reproduced by the recently proposed\cite{Iida09}  simplified model in which 
\begin{table}[h]
  \caption{HOMO and LUMO levels, HOMO-LUMO gap, and excitation energies (EE) in eV using different XCs. }
\small\center
\begin{tabular}{@{}llcccccc}\hline \hline
   & & $E_H$  & $E_L$ & $E_g$  & EE$^{\rm a}$&  EE$^{\rm b}$ & EE-{\rm exp}$^{\rm c}$       \\  \hline
\multirow{4}{*}{Y1} & Gas  & -5.92 & -2.86& 3.06       &  2.93& 3.25           &  -        \\
   & EtOH & -5.91 & -3.00& 2.91    &  2.63& 2.94           &  3.179    \\
   & Dep  & -5.68 & -2.29& 3.39    &  3.14& 3.39           &  -        \\ 
   & Sup   & -5.81&-3.01 & 2.81    &  2.55 &  2.88           &    -      \\ \hline
\multirow{4}{*}{Y2} & Gas  & -6.61 & -3.23& 3.38       &  3.23& 3.48           &  -        \\
   & EtOH & -6.45 & -3.14& 3.31       &  2.97& 3.22           &  3.44     \\
   & Dep  & -5.98 & -2.47& 3.51       & 3.24& 3.46           &  -        \\
   & Sup   & -6.49 & -3.16& 3.33      &  3.03& 3.27           &  -        \\ \hline
   \end{tabular}
	\flushleft $^{\rm a\;\;}${\footnotesize B3LYP} \\
	$^{\rm b\;\;}${\footnotesize CAM-B3LYP} \\
	$^{\rm c\;\;}${\footnotesize Experimental data from Ref.~[\citenum{Han2012}] }
  \label{tab3}
 \end{table}
the level shift was given (within an error of $\sim$0.5~eV for the near HOMO levels) to be proportional to the scalar product of the orbital dipole-moment and the total dipole moment vectors of the molecule in vacuum. The failure is because, (i) the error is larger than the accurate calculated shifts and (ii) looking at dipole moments in Fig.~\ref{fig4} we see that the model predicts all shifts to be downwards (in contradiction to the accurate results listed in Table~\ref{tab3}). Despite this inaccurate prediction of the level shifts, the prediction for the change in HOMO-LUMO gaps are qualitatively correct for our case, and this is because the dipole moments of LUMOs are larger, in magnitude, than those of HOMOs in both Y1 and Y2 molecules. 

Concerning the deprotonated molecules, the results in Table~\ref{tab3} show that deprotonation (in ethanol) of Y1 and Y2 increases the gap by 0.48~eV and 0.20~eV, respectively. This gap-increase may be explained by resorting to a simple picture of quantum potential well: Detachment of a hydrogen nucleus (deprotonation) would give rise to a narrower well and more confined molecular orbitals, which in turn, would lead to upward shift of the levels; the upward shift of the lower-lying HOMO level would be less than that of the LUMO level, and therefore, the level distances (including the HOMO-LUMO gap) would increase. 
\subsubsection{CT-character excited states and solvatochromic shifts}
\begin{figure}[h]
\includegraphics[width=0.9\columnwidth]{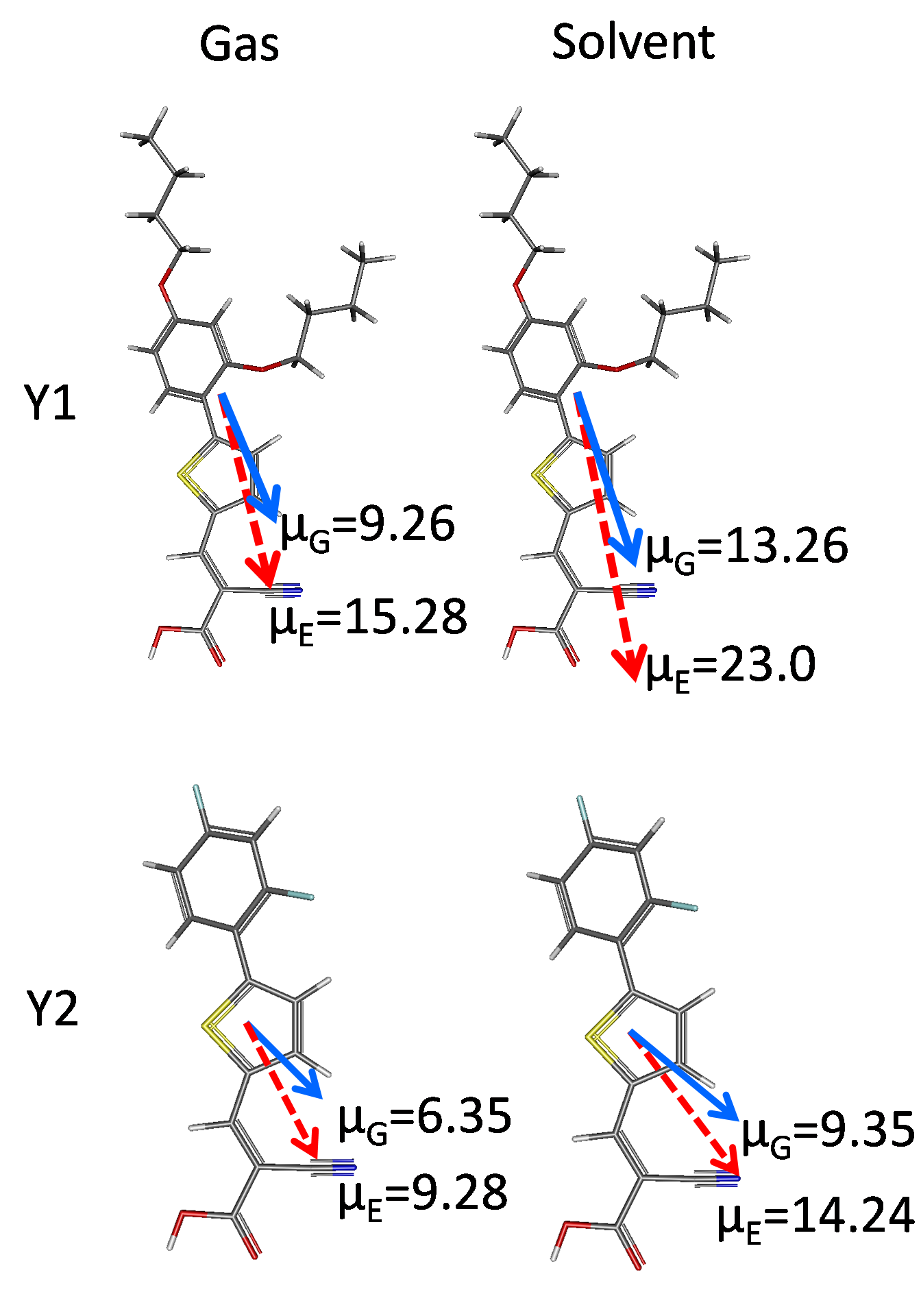}
\caption{Dipole moments of Y1 and Y2 for ground and excited states in both gas phase and solvent. Solid arrows (blue) and dashed arrows (red) correspond to ground and excited states, respectively.}
\label{fig5}
\end{figure}

\begin{figure}[h]
\includegraphics[width=0.9\columnwidth]{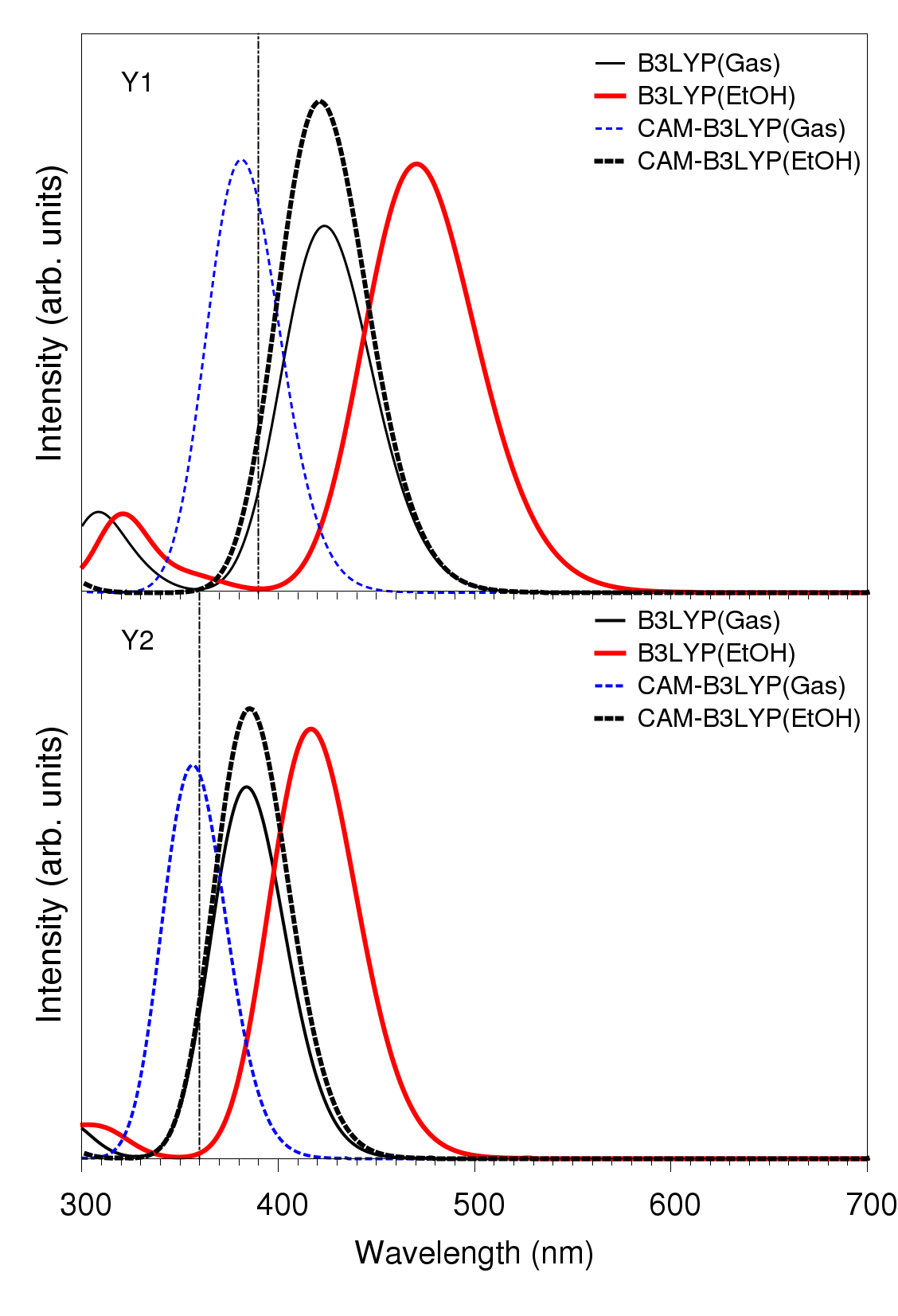}
\caption{UV/Vis absorption spectra for Y1 and Y2 obtained using B3LYP and CAM-B3LYP functionals. Solid and dashed lines correspond to B3LYP and CAM-B3LYP results, respectively. Thin and thick lines represent the gas and solvent phases, respectively. The vertical line indicates the peak position of experiment in ethanol.}
\label{fig6}
\end{figure}

$\;$
The first excitation energies of Y1 and Y2 were calculated using B3LYP and CAM-B3LYP functionals and compared with experiment in Table~\ref{tab3}. Upon moving from the gas phase to the C-PCM solvent description, a red shift of about 0.30 (0.31) eV for Y1 and 0.26 (0.26) eV for Y2 were predicted using B3LYP (CAM-B3LYP) functional. The shift of absorption peak can be related to the dipole moments and reaction fields of the ground and excited states by\cite{Lakowicz06}
\begin{equation}\label{eq4}
\Delta (h\nu)=-\mu_{\rm E}(R_{\rm or}^{\rm G}+R_{\rm el}^{\rm E}) + \mu_{\rm G}(R_{\rm or}^{\rm G}+R_{\rm el}^{\rm G})
\end{equation}
where, $\mu$, $R_{\rm or}$, $R_{\rm el}$ are the molecular dipole moment, reorientational reaction field, electronic reaction field, respectively; and the superscript "E" ("G") denotes the excited (ground) state (See Eqs. 6.9 and 6.10 of Ref.~[\citenum{Lakowicz06}] ). 
Using Eq.~(\ref{eq4}) within the Onsager description\cite{Onsager36} of the reaction field in terms of dipole moments, a red-shift would be observed for $\mu_{\rm E}>\mu_{\rm G}$, and a blue-shift for the reverse case. The L\"owdin charge analysis for the excited states of neutral Y1 and Y2 (see Table~\ref{tab2}), shows that in the first excited state the net charge on the acceptor unit increases and as a result, the corresponding first excited state dipole moments become greater than those of the ground states (see Fig.~\ref{fig5}), and therefore, we end up with red shifts, in agreement with the C-PCM/TDDFT calculation results (see Fig.~\ref{fig6}).

Using the excitation frequencies and oscillator strengths determined from TDDFT calculations, and convoluting them by Eq.~(\ref{eq2})  with $\Delta_{1/2}=0.37$~eV, we have obtained the absorption spectra for Y1 and Y2 in different phases, as shown in Fig.~\ref{fig6}.
As is seen, the first excitation peak is located in the UV/Vis region and higher order excitations (having wavelengths less than $\sim 275$~nm) do not have any contributions in this region.

In neutral and deprotonated configurations of Y1 and Y2, our TDDFT calculations show that in the first excitations, the single-particle HOMO-LUMO transitions have the main contributions compared to other possible transitions, and therefore, the widening of the HOMO-LUMO gap (due to deprotonation) leads to a blue shift in the absorption spectra, as reported in Table~\ref{tab3}.  

To make possible improvements on our C-PCM/TDDFT results, we have repeated the calculations using the solvated supermolecule method, and we did not find any significant deviations from those obtained using C-PCM, which implies that the polarization effects in solvent is dominant compared to the effects of hydrogen bonding between solute and solvent molecules.  

From Table~\ref{tab3}, the excitation energies obtained within CAM-B3LYP shows better agreement with experiment compared to those obtained using B3LYP approximation. The difference between the B3LYP and CAM-B3LYP results is correlated with the extent of the CT-character of the excitations, and this character is quantified by a diagnostic parameter, $\Lambda$, which is defined by\cite{Peach2008}
   
\begin{equation}\label{eq5}
\Lambda=\frac{\sum\kappa_{ia}^2O_{ia}}{\sum\kappa_{ia}^2},
\end{equation}
where $O_{ia}$ is the spatial overlap between occupied $\phi_i$ and virtual $\phi_a$ orbitals, and $\kappa_{ia}$ is the contribution of each occupied-virtual orbital pair in any excitation. For excitations in which the HOMO-LUMO transitions have the dominant contribution, Eq.~(\ref{eq5}) reduces to
\begin{equation}\label{eq6}
\Lambda\approx\int d{\bf r}\;|\phi_L({\bf r})||\phi_H({\bf r})|,
\end{equation}
where $\phi_L$ and $\phi_H$ are the LUMO and HOMO wave functions, respectively. In the present study, we have obtained $\Lambda=0.71$ and 0.74 for Y1 and Y2, respectively. These relatively large values explain the small differences ($\sim 0.3~eV$), reported in Table~\ref{tab3}, between B3LYP and CAM-B3LYP results.

By analyzing the L\"owdin charges of the donor, spacer, and acceptor units of Y1 and Y2 both in ground and excited states, as listed in Table~\ref{tab2}, we can categorize the charge transfers in the excitation processes into two different classes. 
\begin{figure}[h]
\center
\includegraphics[width=0.6\columnwidth]{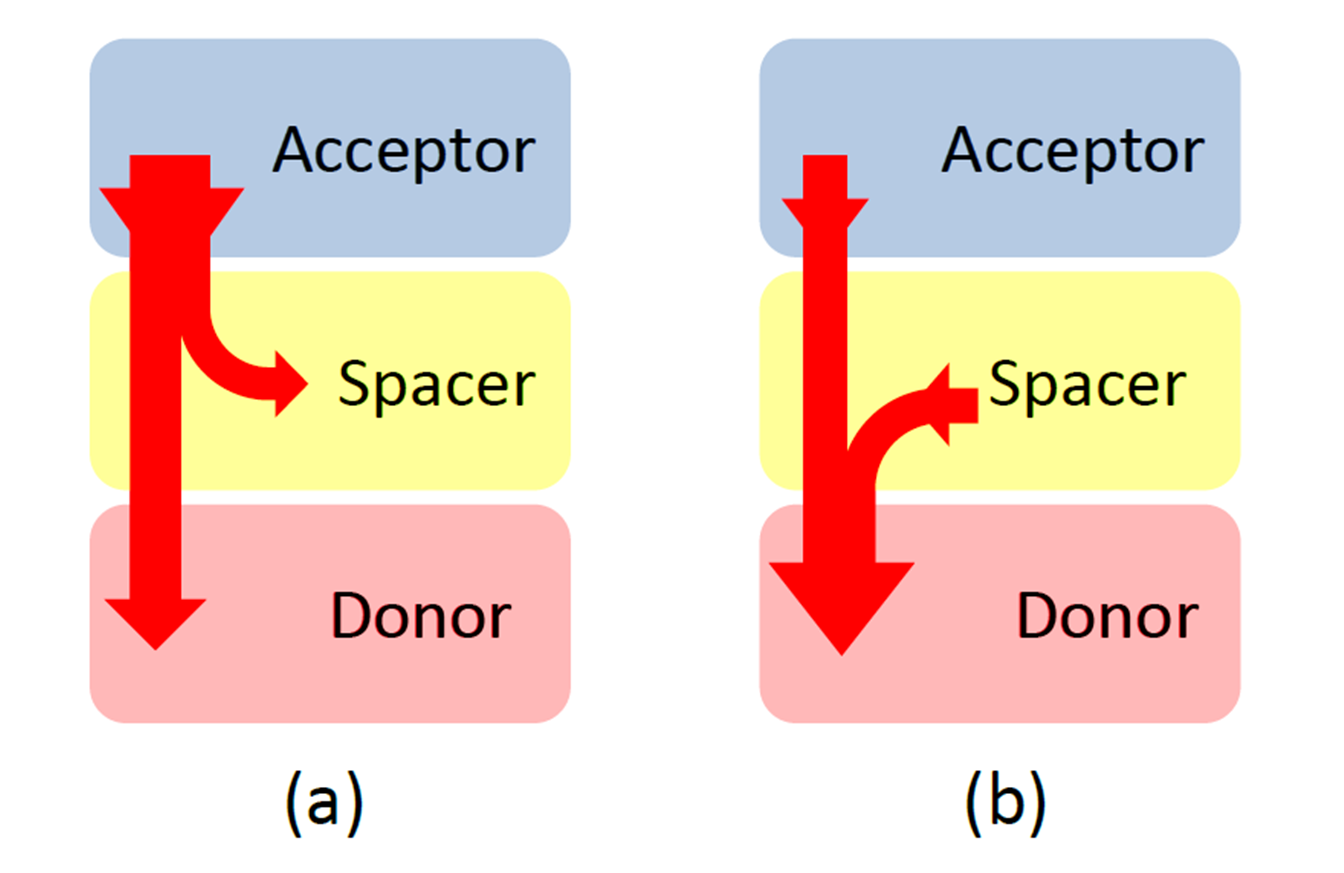}
\caption{Schematic representation of charge-transfer directions in excitation processes. In (a), a net electronic charge is transferred from donor to both spacer and acceptor, describing the excitations of Y1 (except for its deprotonated within CAM-B3LYP); whereas in (b), a net electronic charge is transferred from both donor and spacer to the acceptor, describing the excitations of Y2. Because of the fluorine atoms in Y2, the transferred charge from the donor is smaller than that in Y1.}
\label{fig7}
\end{figure}
Figure~\ref{fig7}(a) schematically represents those excitation processes in which a net electronic charge from the donor is transferred into both spacer and acceptor units, while Fig.~\ref{fig7}(b) represents those excitation processes in which a net electronic charge is transferred from the donor and spacer to the acceptor unit. Excitations of Y1 (except for deprotonated-Y1 within CAM-B3LYP) belong to the former category, whereas the excitations of Y2 belong to the latter category. Because of the fluorine atoms in Y2, the amount of charge transfer from the donor is smaller than that in Y1. For example, from Table~\ref{tab2}, it is -0.27~a.u. for Y1  and -0.11~a.u. for Y2 in solvent within CAM-B3LYP. 

The amount of charge transfers to the acceptor unit during the excitations are obtained from Table~\ref{tab2} and tabulated in Table~\ref{tab4}.        
\begin{table}[h]
\caption{Transferred charge (in a.u.) to the acceptor during the excitation of Y1 and Y2 within CAM-B3LYP. The values in parentheses represent the B3LYP results.}
\small \center
\begin{tabular}{@{}lcccccc}\hline \hline
                      &\; & Gas&\;& Sol&\;& Dep \\ \hline
                     Y1&& -0.17 (-0.18) &&-0.24 (-0.25) &&-0.17 (-0.21)       \\
                     Y2&& -0.13 (-0.13) &&-0.21 (-0.22) &&-0.14 (-0.11)      \\ \hline
 \end{tabular}
 \label{tab4}
\end{table}
 As is seen from Table~\ref{tab4}, in all cases the amount of charge transfer for Y1 is larger than that for Y2. This implies that the amount of charge injection into the TiO$_2$ surface is larger for Y1 than Y2, which in turn, results in a greater short-circuit current density ($J_{\rm sc}$) for Y1, in agreement with the experiment.\cite{Han2012}       
   
\subsection{Electrolyte-Y1/Y2 interactions and $V_{\rm oc}$}
It had been experimentally observed\cite{Han2012} that the open-circuit voltage, $V_{\rm oc}$, for the black dye with coadsorbent Y1 was higher than that with Y2, and it was attributed to the higher recombination time (lower recombination rate) in the coadsorbed N749+Y1 system  compared to N749+Y2. The experiment had also shown that, for N749+Y2, the recombination time is even smaller than that in pure N749 system. 

Some researchers\cite{Zhang2012,Pastore2012} have already proposed that the weak interactions between molecules in the electrolyte solution and the atoms in the dye molecules play the main role in determining the electron lifetime. In other words,  higher concentration of the electrolyte solution molecules near the semiconductor surface (as a result of this interaction) leads to higher probability of electron capture (and therefore lower electron lifetime) from the semiconductor surface. 
It had also been shown\cite{Green2004,Pastore2012} that I$_2$ and I$_3^-$ were the main components of the electrolyte solution which had to be considered, and the rate of electron capture by I$_2$ was two orders of magnitudes greater than that of I$_3^-$. 

To find out the effects of I$_2$ interactions (the I$_3^-$ interactions were ignored) on the recombination time, we benefited the fact that the anchoring carboxylic acid was deposited on TiO$_2$ surface dominantly through bidentate bridging (BB) type adsorption.\cite{Vittadini00,Nazeeruddin2003,Srinivas09,Hara02,Pastore2012chem} We therefore determined, in the first step, the geometry of deposition through BB-type adsorption, using periodic slab model for vacuum and cluster method for solvent calculations. 
\begin{figure}[h]
\includegraphics[width=\columnwidth]{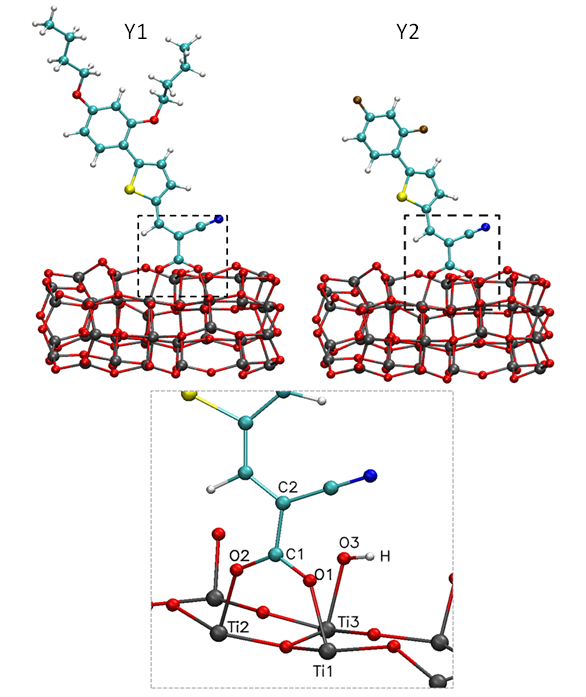}
\caption{Optimized geometry for BB-type adsorption of Y1 and Y2 molecules on (TiO$2$)$_{38}$ cluster in acetonitrile. The dashed square region is zoomed in the bottom.}
\label{fig8}
\end{figure}
The optimized geometries of deposition on (TiO$2$)$_{38}$ cluster surface in acetonitrile is shown in Fig.~\ref{fig8}. As is seen, both Y1 and Y2 molecules were adsorbed almost perpendicularly to the TiO$_2$ surface (more details on adsorption geometry are listed in Table~\ref{tab5}). This relative perpendicular deposition, as we will discuss in the following, plays an important role in the interactions of I$_2$ molecules with the surface. 

\begin{table}[h]
\caption{Interatomic distances, in $\AA$, near the Y1-TiO$_2$ interface and Y2-TiO$_2$ interface both in gas phase and in AN. The molecules are adsorbed on anatase TiO$_2$ (101) surface. The labels are defined in the bottom sub-figure of Fig.~\ref{fig8}. The angle between O$_1$C$_1$ and C$_1$O$_2$ bonds is equal to 125 degrees in all cases.}
\small 
\begin{tabular}{@{}llcccccc}\hline \hline
                    &  & Ti$_1$O$_1$& Ti$_2$O$_2$& C$_1$O$_1$& C$_1$O$_2$& C$_1$C$_2$ &  O$_3$H \\ \hline
\multirow{2}{*}{Y1} & vac & 2.07& 2.09 & 1.28&  1.28 & 1.48 &  0.98     \\
                    & AN  & 2.08& 2.07 & 1.27&  1.28 & 1.48 &  0.97     \\ \hline
\multirow{2}{*}{Y2} & vac & 2.08& 2.09 & 1.28&  1.27 & 1.48 &  0.98     \\
                    & AN  & 2.11& 2.09 & 1.28&  1.27 & 1.48 &  0.98     \\  \hline
 \end{tabular}
 \label{tab5}
\end{table}

\begin{figure}[h]\center
\includegraphics[width=0.9\columnwidth]{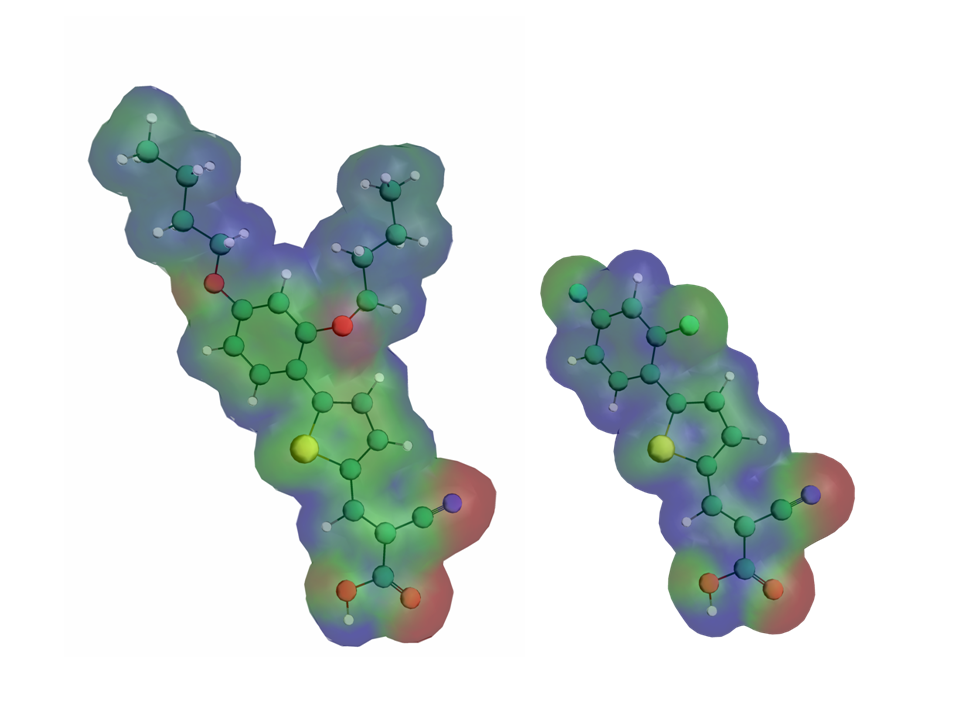}
\caption{Calculated molecular electrostatic potential on the 0.005 a.u. isodensity surface of standalone Y1 and Y2 molecules. The color range from red to blue identifies the most negative to most positive potentials.     }
\label{fig9}
\end{figure}

In the second step, we explored the most attractive sites on standalone Y1 and Y2 molecules by plotting the molecular electrostatic potential (MEP) maps on a 0.005 a.u. isodensity surface (Fig.~\ref{fig9}). 
Analyzing these maps,\cite{Politzer07,Politzer10,Politzer13} the CN groups (red color) were identified as the most attractive sites on the molecules to form halogen bonding with I$_2$ through $\sigma$-hole interaction. The oxygen, fluorine, and sulfur atoms were identified as other (weaker) attractive sites. To determine the geometry and interaction energies of I$_2$ molecules with these attractive sites, we have performed DFT calculations using NWChem within 6-311G**/B3LYP level of approximation. It should be mentioned that the accuracy of ordinary DFT calculations for these weak intermolecular interactions had already been demonstrated by others.\cite{Pastore2012}          
\begin{figure}[h]\center
\includegraphics[width=0.8\columnwidth]{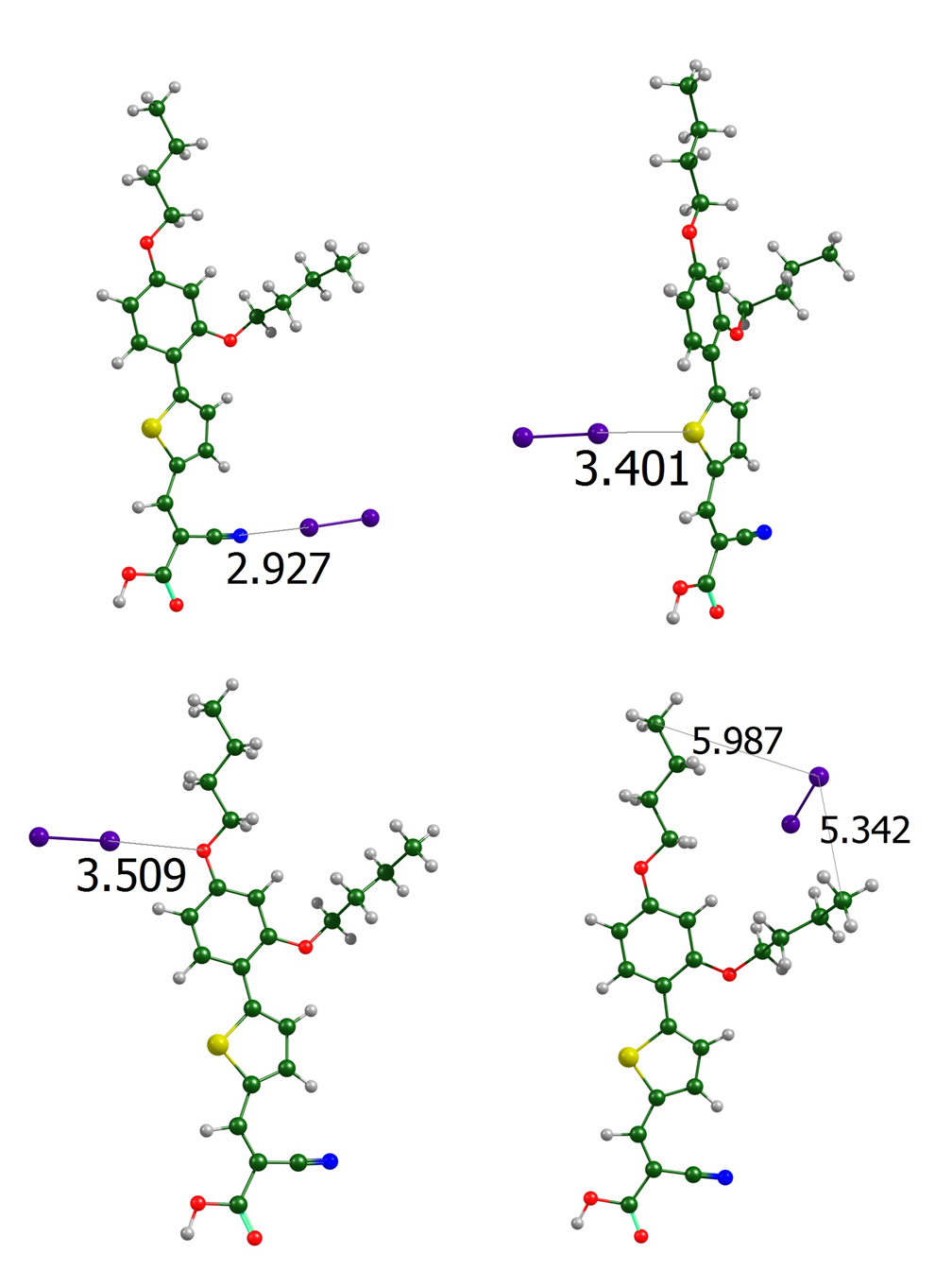}
\caption{Optimized structures of Y1-I$_2$ adducts with the relative distances, in $\AA$, of the I$_2$ molecules from Y1.}
\label{fig10}
\end{figure}
\begin{figure}[h]\center
\includegraphics[width=0.5\columnwidth]{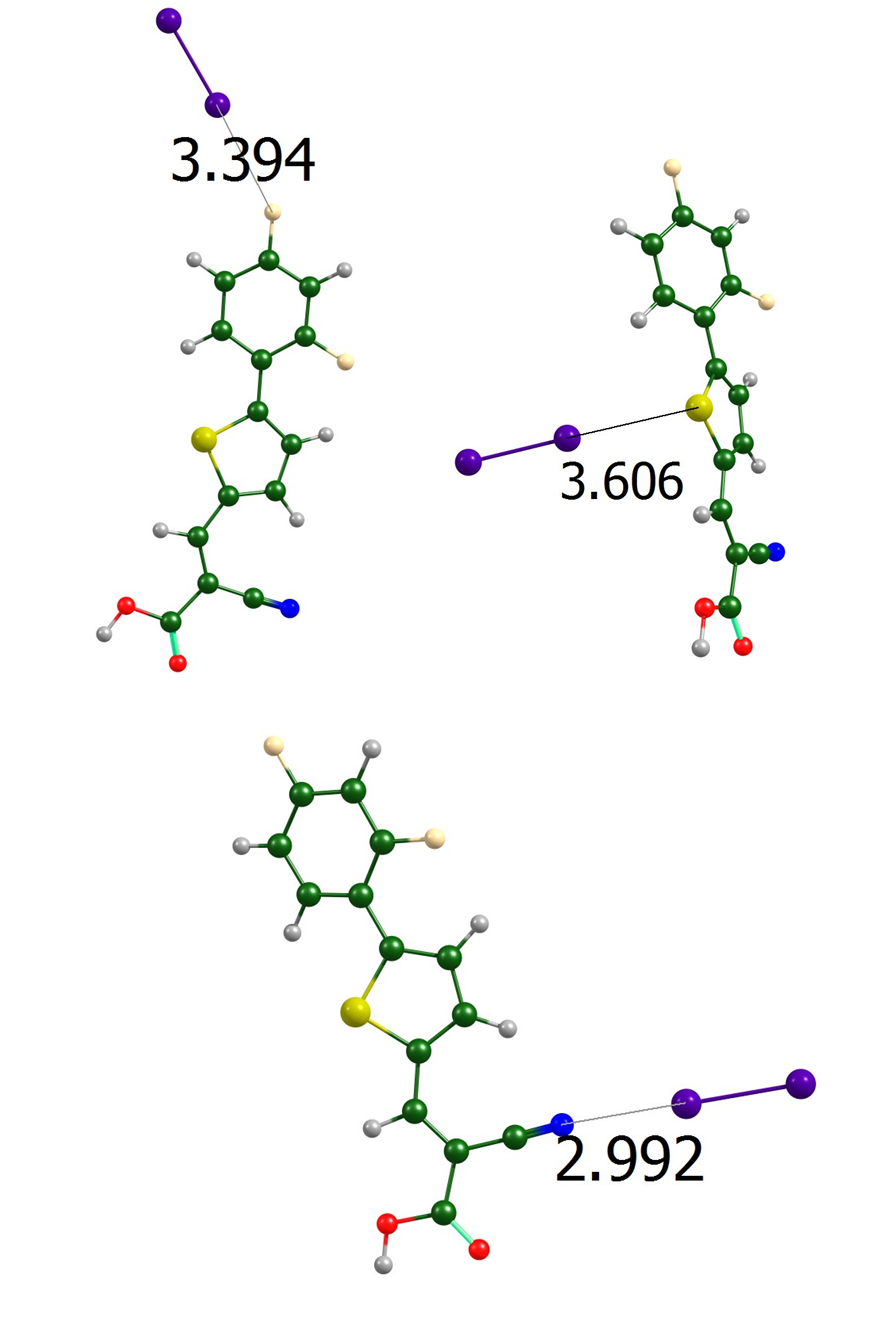}
\caption{The same as in Fig.~\ref{fig10} for Y2.}
\label{fig11}
\end{figure}

The optimized geometries of the Y1-I$_2$ and Y2-I$_2$ adducts were determined for different attractive sites (S, N, O, and F atoms) on the standalone molecules and shown in Fig.~\ref{fig10} and Fig.~\ref{fig11}, respectively. The geometries clearly show the directional character of the halogen bonding.\cite{Politzer10} The binding energies of the adducts and corresponding distances are listed in Table~\ref{tab6} for cases with and without basis-set-superposition-error (BSSE) corrections.\cite{boys1970} 
\begin{table}[h]
  \caption{The BSSE corrected equilibrium distances in $\AA$, and interaction energies in kcal/mol of I$_2$ with Y1 and Y2 for different interaction sites. The values in parentheses correspond to uncorrected quantities.}
\center
\begin{tabular}{@{}lllccc} \hline \hline
            &&Site &&$d$ ($\AA$)& $-E_{\rm int}$ (kcal/mol)  \\ \hline
\multirow{4}{*}{Y1}& & I$_2$-N && 2.93 (2.89) & 4.38 (5.41)   \\
                   & & I$_2$-S&& 3.40 (3.27) & 0.86 (1.10)  \\ 
                   & & I$_2$-O& &3.51 (3.37) & 0.26 (0.98)     \\ \hline
\multirow{3}{*}{Y2} && I$_2$-N && 2.99 (2.96) & 3.79 (4.56)     \\
                    && I$_2$-S&& 3.61 (3.46) & 0.28 (1.01)     \\ 
                    &&I$_2$-F && 3.39 (3.22) & 0.55 (1.36)   \\ \hline    	        
\end{tabular}
\label{tab6}
\end{table}
These results show that, without BSSE correction the calculations give stronger bonds with shorter halogen bond lengths. Moreover, it has been shown\cite{Salvador01} that for weaker bonds the BSSE correction has significant effects on the geometries. 
Our calculated binding energies show that the CN groups are the most preferred sites to hold I$_2$ molecules, as predicted before from the MEP maps in Fig.~\ref{fig9}. 

Taking into account the geometry of perpendicular Y1 and Y2 depositions on the surface, if I$_2$ molecules were to access these sites, they would pass through the top side of the coadsorbents. In Y1 case, we obtained that the I$_2$ molecules stay at optimized distances of 5.34 to 5.99~$\AA$ from the topmost carbon atoms of butyloxyl groups (as shown in Fig.~\ref{fig10}) with very low binding energies ($\sim 0.03$ kcal/mol). This result implies that the butyloxyl groups in Y1 hinder the accumulation of I$_2$ molecules near the attractive sites (and thereby near the surface). However, there is no such blocker in Y2. The overall effect is that, when Y1 is coadsorbed with black dye, the iodine concentration near the surface decreases compared to pure black dye, whereas for Y2 the iodine concentration increases compared to pure black dye. As a result, coadsorbing Y1 with black dye increases the recombination time (and therebye increases $V_{oc}$) while coadsorbing Y2 decreases it.      

As we discussed, both $J_{sc}$ and $V_{oc}$ in case of coadsorbing Y1 are greater than those when Y2 is coadsorbed, and the product of them, which is proportional to the efficiency of the cell, is also greater in Y1 case. This proves the higher efficiency of the black dye cell when coadsorbed with Y1, in agreement with experiment.\cite{Han2012}

\section{Conclusions}\label{sec4}
In this work, we have studied the effects of coadsorbing Y1 and Y2 on the efficiency of black dye sensitized solar cell, and have theoretically explained why Y1 would increase the efficiency. We started by determining the equilibrium geometries of the coadsorbents in different media using C-PCM for solvents. The results showed that thiophene and cyanoacetic groups were coplanar, giving rise to a strong conjugation across the acceptor-spacer groups. Using the more sophisticated "solvated supermolecule model", only the OH bond length of carboxylic group changed from the C-PCM value 0.97 to 1.01~$\AA$ with no significant changes in other parameters. In optical properties calculations within TDDFT, the redshifts in polar solvents and the blue shifts due to deprotonation were correlated with the decrease and increase of the HOMO-LUMO gap, respectively. By calculating the ground- and excited-state dipole moments, we have also obtained the solvatochromic redshift, in agreement with the TDDFT results, and also showed that the solvatochromic shift would be a redshift for $\mu_{\rm E}>\mu_{G}$, and a blueshift for $\mu_{\rm E}<\mu_{G}$. The charge density analysis for the ground and excited states revealed that in the excitation process the charge transfer to the acceptor is larger in Y1 than Y2, which explains the larger $J_{sc}$ for Y1. Concerning $V_{oc}$, we considered the interactions of I$_2$ molecules with the coadsorbents and found out that in Y1, because of the butyloxyl groups, the accumulation of I$_2$ near the surface is hindered, leading to a higher recombination time, and thereby a larger $V_{oc}$. On the other hand, in Y2, the I$_2$ concentration near the surface is increased due to the interactions with the CN group, leading to a lower recombination time. The overall result was that coadsorbing Y1 with a black dye would increase the cell efficiency. It should be emphasized that charge analysis for the excited states helped much in determining the behavior of $J_{sc}$, and MEP maps helped to identify the attractive sites on the coadsorbents. These tools will be utilized in designing newer dye and coadsorbent molecules with higher efficiencies.            
\section*{Acknowledgement} 
Y.~T.~A. would like to thank Professor F.~De Angelis for the discussion on constructing the TiO$_2$ cluster. This work is part of research program in Theoretical and Computational Physics Group, AEOI.  

\footnotesize{
\bibliography{revised-payami} 

\providecommand*{\mcitethebibliography}{\thebibliography}
\csname @ifundefined\endcsname{endmcitethebibliography}
{\let\endmcitethebibliography\endthebibliography}{}
\begin{mcitethebibliography}{51}
\providecommand*{\natexlab}[1]{#1}
\providecommand*{\mciteSetBstSublistMode}[1]{}
\providecommand*{\mciteSetBstMaxWidthForm}[2]{}
\providecommand*{\mciteBstWouldAddEndPuncttrue}
  {\def\EndOfBibitem{\unskip.}}
\providecommand*{\mciteBstWouldAddEndPunctfalse}
  {\let\EndOfBibitem\relax}
\providecommand*{\mciteSetBstMidEndSepPunct}[3]{}
\providecommand*{\mciteSetBstSublistLabelBeginEnd}[3]{}
\providecommand*{\EndOfBibitem}{}
\mciteSetBstSublistMode{f}
\mciteSetBstMaxWidthForm{subitem}
{(\emph{\alph{mcitesubitemcount}})}
\mciteSetBstSublistLabelBeginEnd{\mcitemaxwidthsubitemform\space}
{\relax}{\relax}

\bibitem[O'Regan and Gratzel(1991)]{Oregan1991}
B.~O'Regan and M.~Gratzel, \emph{nature}, 1991, \textbf{353}, 737--740\relax
\mciteBstWouldAddEndPuncttrue
\mciteSetBstMidEndSepPunct{\mcitedefaultmidpunct}
{\mcitedefaultendpunct}{\mcitedefaultseppunct}\relax
\EndOfBibitem
\bibitem[Hagfeldt \emph{et~al.}(2010)Hagfeldt, Boschloo, Sun, Kloo, and
  Pettersson]{hagfeldt2010}
A.~Hagfeldt, G.~Boschloo, L.~Sun, L.~Kloo and H.~Pettersson, \emph{Chemical
  Reviews}, 2010, \textbf{110}, 6595--6663\relax
\mciteBstWouldAddEndPuncttrue
\mciteSetBstMidEndSepPunct{\mcitedefaultmidpunct}
{\mcitedefaultendpunct}{\mcitedefaultseppunct}\relax
\EndOfBibitem
\bibitem[O'Regan and Durrant(2009)]{oregan2009}
B.~C. O'Regan and J.~Durrant, \emph{Accounts of chemical research}, 2009,
  \textbf{42}, 1799--1808\relax
\mciteBstWouldAddEndPuncttrue
\mciteSetBstMidEndSepPunct{\mcitedefaultmidpunct}
{\mcitedefaultendpunct}{\mcitedefaultseppunct}\relax
\EndOfBibitem
\bibitem[Palomares \emph{et~al.}(2002)Palomares, Clifford, Haque, Lutz, and
  Durrant]{Pal02}
E.~Palomares, J.~N. Clifford, S.~A. Haque, T.~Lutz and J.~R. Durrant,
  \emph{Journal of the American Chemical Society}, 2002, \textbf{125},
  475--482\relax
\mciteBstWouldAddEndPuncttrue
\mciteSetBstMidEndSepPunct{\mcitedefaultmidpunct}
{\mcitedefaultendpunct}{\mcitedefaultseppunct}\relax
\EndOfBibitem
\bibitem[O'Regan \emph{et~al.}(2008)O'Regan, López-Duarte, Martínez-Díaz,
  Forneli, Albero, Morandeira, Palomares, Torres, and Durrant]{oregan2008}
B.~C. O'Regan, I.~López-Duarte, M.~V. Martínez-Díaz, A.~Forneli, J.~Albero,
  A.~Morandeira, E.~Palomares, T.~Torres and J.~R. Durrant, \emph{Journal of
  the American Chemical Society}, 2008, \textbf{130}, 2906--2907\relax
\mciteBstWouldAddEndPuncttrue
\mciteSetBstMidEndSepPunct{\mcitedefaultmidpunct}
{\mcitedefaultendpunct}{\mcitedefaultseppunct}\relax
\EndOfBibitem
\bibitem[Maggio \emph{et~al.}(2013)Maggio, Martsinovich, and
  Troisi]{Maggio2013}
E.~Maggio, N.~Martsinovich and A.~Troisi, \emph{Angewandte Chemie International
  Edition}, 2013, \textbf{52}, 973--975\relax
\mciteBstWouldAddEndPuncttrue
\mciteSetBstMidEndSepPunct{\mcitedefaultmidpunct}
{\mcitedefaultendpunct}{\mcitedefaultseppunct}\relax
\EndOfBibitem
\bibitem[Lee \emph{et~al.}(2012)Lee, Park, Cho, Son, Sudhagar, Jung, Wooh,
  Char, and Kang]{Lee2012}
Y.-G. Lee, S.~Park, W.~Cho, T.~Son, P.~Sudhagar, J.~H. Jung, S.~Wooh, K.~Char
  and Y.~S. Kang, \emph{The Journal of Physical Chemistry C}, 2012,
  \textbf{116}, 6770--6777\relax
\mciteBstWouldAddEndPuncttrue
\mciteSetBstMidEndSepPunct{\mcitedefaultmidpunct}
{\mcitedefaultendpunct}{\mcitedefaultseppunct}\relax
\EndOfBibitem
\bibitem[Zhang \emph{et~al.}(2012)Zhang, Kan, Li, Geng, Wu, and Su]{Zhang2012}
J.~Zhang, Y.-H. Kan, H.-B. Li, Y.~Geng, Y.~Wu and Z.-M. Su, \emph{Dyes and
  Pigments}, 2012, \textbf{95}, 313--321\relax
\mciteBstWouldAddEndPuncttrue
\mciteSetBstMidEndSepPunct{\mcitedefaultmidpunct}
{\mcitedefaultendpunct}{\mcitedefaultseppunct}\relax
\EndOfBibitem
\bibitem[Boschloo and Hagfeldt(2009)]{Boschloo2009}
G.~Boschloo and A.~Hagfeldt, \emph{Accounts of Chemical Research}, 2009,
  \textbf{42}, 1819--1826\relax
\mciteBstWouldAddEndPuncttrue
\mciteSetBstMidEndSepPunct{\mcitedefaultmidpunct}
{\mcitedefaultendpunct}{\mcitedefaultseppunct}\relax
\EndOfBibitem
\bibitem[Wang \emph{et~al.}(2003)Wang, Zakeeruddin, Moser, and
  Grätzel]{Wang2003}
P.~Wang, S.~M. Zakeeruddin, J.-E. Moser and M.~Grätzel, \emph{The Journal of
  Physical Chemistry B}, 2003, \textbf{107}, 13280--13285\relax
\mciteBstWouldAddEndPuncttrue
\mciteSetBstMidEndSepPunct{\mcitedefaultmidpunct}
{\mcitedefaultendpunct}{\mcitedefaultseppunct}\relax
\EndOfBibitem
\bibitem[Han \emph{et~al.}(2012)Han, Islam, Chen, Malapaka, Chiranjeevi, Zhang,
  Yang, and Yanagida]{Han2012}
L.~Han, A.~Islam, H.~Chen, C.~Malapaka, B.~Chiranjeevi, S.~Zhang, X.~Yang and
  M.~Yanagida, \emph{Energy and Environmental Science}, 2012, \textbf{5},
  6057--6060\relax
\mciteBstWouldAddEndPuncttrue
\mciteSetBstMidEndSepPunct{\mcitedefaultmidpunct}
{\mcitedefaultendpunct}{\mcitedefaultseppunct}\relax
\EndOfBibitem
\bibitem[Hohenberg and Kohn(1964)]{HK64}
P.~Hohenberg and W.~Kohn, \emph{Phys. Rev.}, 1964, \textbf{136},
  B864--B871\relax
\mciteBstWouldAddEndPuncttrue
\mciteSetBstMidEndSepPunct{\mcitedefaultmidpunct}
{\mcitedefaultendpunct}{\mcitedefaultseppunct}\relax
\EndOfBibitem
\bibitem[Runge and Gross(1984)]{RG84}
E.~Runge and E.~K.~U. Gross, \emph{Phys. Rev. Lett.}, 1984, \textbf{52},
  997--1000\relax
\mciteBstWouldAddEndPuncttrue
\mciteSetBstMidEndSepPunct{\mcitedefaultmidpunct}
{\mcitedefaultendpunct}{\mcitedefaultseppunct}\relax
\EndOfBibitem
\bibitem[Becke(1993)]{B3LYP93-1}
A.~D. Becke, \emph{The Journal of Chemical Physics}, 1993, \textbf{98},
  1372--1377\relax
\mciteBstWouldAddEndPuncttrue
\mciteSetBstMidEndSepPunct{\mcitedefaultmidpunct}
{\mcitedefaultendpunct}{\mcitedefaultseppunct}\relax
\EndOfBibitem
\bibitem[Becke(1993)]{B3LYP93-2}
A.~D. Becke, \emph{The Journal of Chemical Physics}, 1993, \textbf{98},
  5648--5652\relax
\mciteBstWouldAddEndPuncttrue
\mciteSetBstMidEndSepPunct{\mcitedefaultmidpunct}
{\mcitedefaultendpunct}{\mcitedefaultseppunct}\relax
\EndOfBibitem
\bibitem[Yanai \emph{et~al.}(2004)Yanai, Tew, and Handy]{Yanai2004}
T.~Yanai, D.~P. Tew and N.~C. Handy, \emph{Chemical Physics Letters}, 2004,
  \textbf{393}, 51--57\relax
\mciteBstWouldAddEndPuncttrue
\mciteSetBstMidEndSepPunct{\mcitedefaultmidpunct}
{\mcitedefaultendpunct}{\mcitedefaultseppunct}\relax
\EndOfBibitem
\bibitem[Maggio \emph{et~al.}(2012)Maggio, Martsinovich, and
  Troisi]{Troisi2012}
E.~Maggio, N.~Martsinovich and A.~Troisi, \emph{The Journal of Physical
  Chemistry C}, 2012, \textbf{116}, 7638--7649\relax
\mciteBstWouldAddEndPuncttrue
\mciteSetBstMidEndSepPunct{\mcitedefaultmidpunct}
{\mcitedefaultendpunct}{\mcitedefaultseppunct}\relax
\EndOfBibitem
\bibitem[Schmidt \emph{et~al.}(1993)Schmidt, Baldridge, Boatz, Elbert, Gordon,
  Jensen, Koseki, Matsunaga, Nguyen, and Su]{Schmidt1993}
M.~W. Schmidt, K.~K. Baldridge, J.~A. Boatz, S.~T. Elbert, M.~S. Gordon, J.~H.
  Jensen, S.~Koseki, N.~Matsunaga, K.~A. Nguyen and S.~Su, \emph{Journal of
  Computational Chemistry}, 1993, \textbf{14}, 1347--1363\relax
\mciteBstWouldAddEndPuncttrue
\mciteSetBstMidEndSepPunct{\mcitedefaultmidpunct}
{\mcitedefaultendpunct}{\mcitedefaultseppunct}\relax
\EndOfBibitem
\bibitem[Cramer and Truhlar(1999)]{Cramer1999}
C.~J. Cramer and D.~G. Truhlar, \emph{Chemical Reviews}, 1999, \textbf{99},
  2161--2200\relax
\mciteBstWouldAddEndPuncttrue
\mciteSetBstMidEndSepPunct{\mcitedefaultmidpunct}
{\mcitedefaultendpunct}{\mcitedefaultseppunct}\relax
\EndOfBibitem
\bibitem[Tomasi \emph{et~al.}(2005)Tomasi, Mennucci, and Cammi]{Tomasi2005}
J.~Tomasi, B.~Mennucci and R.~Cammi, \emph{Chemical Reviews-Columbus}, 2005,
  \textbf{105}, 2999--3094\relax
\mciteBstWouldAddEndPuncttrue
\mciteSetBstMidEndSepPunct{\mcitedefaultmidpunct}
{\mcitedefaultendpunct}{\mcitedefaultseppunct}\relax
\EndOfBibitem
\bibitem[Klamt and Sch\"u\"urmann(1993)]{Klamt93}
A.~Klamt and G.~Sch\"u\"urmann, \emph{J. Chem. Soc.{,} Perkin Trans. 2}, 1993,
  799--805\relax
\mciteBstWouldAddEndPuncttrue
\mciteSetBstMidEndSepPunct{\mcitedefaultmidpunct}
{\mcitedefaultendpunct}{\mcitedefaultseppunct}\relax
\EndOfBibitem
\bibitem[Cappelli \emph{et~al.}(2000)Cappelli, Mennucci, da~Silva, and
  Tomasi]{Cappelli2000}
C.~Cappelli, B.~Mennucci, C.~O. da~Silva and J.~Tomasi, \emph{The Journal of
  Chemical Physics}, 2000, \textbf{112}, 5382--5392\relax
\mciteBstWouldAddEndPuncttrue
\mciteSetBstMidEndSepPunct{\mcitedefaultmidpunct}
{\mcitedefaultendpunct}{\mcitedefaultseppunct}\relax
\EndOfBibitem
\bibitem[Marenich \emph{et~al.}(2011)Marenich, Cramer, Truhlar, Guido,
  Mennucci, Scalmani, and Frisch]{Marenich2011}
A.~V. Marenich, C.~J. Cramer, D.~G. Truhlar, C.~A. Guido, B.~Mennucci,
  G.~Scalmani and M.~J. Frisch, \emph{Chemical Science}, 2011, \textbf{2},
  2143--2161\relax
\mciteBstWouldAddEndPuncttrue
\mciteSetBstMidEndSepPunct{\mcitedefaultmidpunct}
{\mcitedefaultendpunct}{\mcitedefaultseppunct}\relax
\EndOfBibitem
\bibitem[Casida(1995)]{Casida1995}
M.~E. Casida, \emph{Time-dependent density functional response theory for
  molecules}, World Scientific: Singapore, 1995, vol.~1\relax
\mciteBstWouldAddEndPuncttrue
\mciteSetBstMidEndSepPunct{\mcitedefaultmidpunct}
{\mcitedefaultendpunct}{\mcitedefaultseppunct}\relax
\EndOfBibitem
\bibitem[Dreuw and Head-Gordon(2005)]{Dreuw2005}
A.~Dreuw and M.~Head-Gordon, \emph{Chemical Reviews}, 2005, \textbf{105},
  4009--4037\relax
\mciteBstWouldAddEndPuncttrue
\mciteSetBstMidEndSepPunct{\mcitedefaultmidpunct}
{\mcitedefaultendpunct}{\mcitedefaultseppunct}\relax
\EndOfBibitem
\bibitem[Casida(2009)]{Casida2009}
M.~E. Casida, \emph{Journal of Molecular Structure: THEOCHEM}, 2009,
  \textbf{914}, 3--18\relax
\mciteBstWouldAddEndPuncttrue
\mciteSetBstMidEndSepPunct{\mcitedefaultmidpunct}
{\mcitedefaultendpunct}{\mcitedefaultseppunct}\relax
\EndOfBibitem
\bibitem[Dreuw and Head-Gordon(2004)]{Dreuw2004}
A.~Dreuw and M.~Head-Gordon, \emph{Journal of the American Chemical Society},
  2004, \textbf{126}, 4007--4016\relax
\mciteBstWouldAddEndPuncttrue
\mciteSetBstMidEndSepPunct{\mcitedefaultmidpunct}
{\mcitedefaultendpunct}{\mcitedefaultseppunct}\relax
\EndOfBibitem
\bibitem[Adamo and Jacquemin(2013)]{adamo2013}
C.~Adamo and D.~Jacquemin, \emph{Chemical Society Reviews}, 2013, \textbf{42},
  845--856\relax
\mciteBstWouldAddEndPuncttrue
\mciteSetBstMidEndSepPunct{\mcitedefaultmidpunct}
{\mcitedefaultendpunct}{\mcitedefaultseppunct}\relax
\EndOfBibitem
\bibitem[Cossi and Barone(2001)]{Cossi2001}
M.~Cossi and V.~Barone, \emph{The Journal of chemical physics}, 2001,
  \textbf{115}, 4708\relax
\mciteBstWouldAddEndPuncttrue
\mciteSetBstMidEndSepPunct{\mcitedefaultmidpunct}
{\mcitedefaultendpunct}{\mcitedefaultseppunct}\relax
\EndOfBibitem
\bibitem[Kohn and Sham(1965)]{KS65}
W.~Kohn and L.~J. Sham, \emph{Physical Review}, 1965, \textbf{140}, A1133\relax
\mciteBstWouldAddEndPuncttrue
\mciteSetBstMidEndSepPunct{\mcitedefaultmidpunct}
{\mcitedefaultendpunct}{\mcitedefaultseppunct}\relax
\EndOfBibitem
\bibitem[Giannozzi \emph{et~al.}(2009)Giannozzi, Baroni, Bonini, Calandra, Car,
  Cavazzoni, Ceresoli, Chiarotti, Cococcioni, Dabo, {Dal Corso}, de~Gironcoli,
  Fabris, Fratesi, Gebauer, Gerstmann, Gougoussis, Kokalj, Lazzeri,
  Martin-Samos, Marzari, Mauri, Mazzarello, Paolini, Pasquarello, Paulatto,
  Sbraccia, Scandolo, Sclauzero, Seitsonen, Smogunov, Umari, and
  Wentzcovitch]{QE-2009}
P.~Giannozzi, S.~Baroni, N.~Bonini, M.~Calandra, R.~Car, C.~Cavazzoni,
  D.~Ceresoli, G.~L. Chiarotti, M.~Cococcioni, I.~Dabo, A.~{Dal Corso},
  S.~de~Gironcoli, S.~Fabris, G.~Fratesi, R.~Gebauer, U.~Gerstmann,
  C.~Gougoussis, A.~Kokalj, M.~Lazzeri, L.~Martin-Samos, N.~Marzari, F.~Mauri,
  R.~Mazzarello, S.~Paolini, A.~Pasquarello, L.~Paulatto, C.~Sbraccia,
  S.~Scandolo, G.~Sclauzero, A.~P. Seitsonen, A.~Smogunov, P.~Umari and R.~M.
  Wentzcovitch, \emph{Journal of Physics: Condensed Matter}, 2009, \textbf{21},
  395502 (19pp)\relax
\mciteBstWouldAddEndPuncttrue
\mciteSetBstMidEndSepPunct{\mcitedefaultmidpunct}
{\mcitedefaultendpunct}{\mcitedefaultseppunct}\relax
\EndOfBibitem
\bibitem[Perdew \emph{et~al.}(1996)Perdew, Burke, and Ernzerhof]{PBE96}
J.~P. Perdew, K.~Burke and M.~Ernzerhof, \emph{Phys. Rev. Lett.}, 1996,
  \textbf{77}, 3865--3868\relax
\mciteBstWouldAddEndPuncttrue
\mciteSetBstMidEndSepPunct{\mcitedefaultmidpunct}
{\mcitedefaultendpunct}{\mcitedefaultseppunct}\relax
\EndOfBibitem
\bibitem[Persson \emph{et~al.}(2000)Persson, Bergström, and
  Lunell]{Persson2000}
P.~Persson, R.~Bergström and S.~Lunell, \emph{The Journal of Physical
  Chemistry B}, 2000, \textbf{104}, 10348--10351\relax
\mciteBstWouldAddEndPuncttrue
\mciteSetBstMidEndSepPunct{\mcitedefaultmidpunct}
{\mcitedefaultendpunct}{\mcitedefaultseppunct}\relax
\EndOfBibitem
\bibitem[Pastore and De~Angelis(2012)]{Pastore2012chem}
M.~Pastore and F.~De~Angelis, \emph{Physical Chemistry Chemical Physics}, 2012,
  \textbf{14}, 920--928\relax
\mciteBstWouldAddEndPuncttrue
\mciteSetBstMidEndSepPunct{\mcitedefaultmidpunct}
{\mcitedefaultendpunct}{\mcitedefaultseppunct}\relax
\EndOfBibitem
\bibitem[De~Angelis \emph{et~al.}(2008)De~Angelis, Fantacci, and
  Selloni]{deangelis2008}
F.~De~Angelis, S.~Fantacci and A.~Selloni, \emph{Nanotechnology}, 2008,
  \textbf{19}, 424002\relax
\mciteBstWouldAddEndPuncttrue
\mciteSetBstMidEndSepPunct{\mcitedefaultmidpunct}
{\mcitedefaultendpunct}{\mcitedefaultseppunct}\relax
\EndOfBibitem
\bibitem[Valiev \emph{et~al.}(2010)Valiev, Bylaska, Govind, Kowalski,
  Straatsma, Van~Dam, Wang, Nieplocha, Apra, Windus, and de~Jong]{Valiev2010}
M.~Valiev, E.~J. Bylaska, N.~Govind, K.~Kowalski, T.~P. Straatsma, H.~J.~J.
  Van~Dam, D.~Wang, J.~Nieplocha, E.~Apra, T.~L. Windus and W.~A. de~Jong,
  \emph{Computer Physics Communications}, 2010, \textbf{181}, 1477--1489\relax
\mciteBstWouldAddEndPuncttrue
\mciteSetBstMidEndSepPunct{\mcitedefaultmidpunct}
{\mcitedefaultendpunct}{\mcitedefaultseppunct}\relax
\EndOfBibitem
\bibitem[Boys and Bernardi(1970)]{boys1970}
S.~F. Boys and F.~d. Bernardi, \emph{Molecular Physics}, 1970, \textbf{19},
  553--566\relax
\mciteBstWouldAddEndPuncttrue
\mciteSetBstMidEndSepPunct{\mcitedefaultmidpunct}
{\mcitedefaultendpunct}{\mcitedefaultseppunct}\relax
\EndOfBibitem
\bibitem[Iida \emph{et~al.}(2009)Iida, Yokogawa, Sato, and Sakaki]{Iida09}
K.~Iida, D.~Yokogawa, H.~Sato and S.~Sakaki, \emph{The Journal of chemical
  physics}, 2009, \textbf{130}, 044107\relax
\mciteBstWouldAddEndPuncttrue
\mciteSetBstMidEndSepPunct{\mcitedefaultmidpunct}
{\mcitedefaultendpunct}{\mcitedefaultseppunct}\relax
\EndOfBibitem
\bibitem[Lakowicz(2006)]{Lakowicz06}
J.~R. Lakowicz, \emph{Principles of Fluorescence Spectroscopy}, Springer, 3rd
  edn, 2006\relax
\mciteBstWouldAddEndPuncttrue
\mciteSetBstMidEndSepPunct{\mcitedefaultmidpunct}
{\mcitedefaultendpunct}{\mcitedefaultseppunct}\relax
\EndOfBibitem
\bibitem[Onsager(1936)]{Onsager36}
L.~Onsager, \emph{Journal of the American Chemical Society}, 1936, \textbf{58},
  1486--1493\relax
\mciteBstWouldAddEndPuncttrue
\mciteSetBstMidEndSepPunct{\mcitedefaultmidpunct}
{\mcitedefaultendpunct}{\mcitedefaultseppunct}\relax
\EndOfBibitem
\bibitem[Peach \emph{et~al.}(2008)Peach, Benfield, Helgaker, and
  Tozer]{Peach2008}
M.~J. Peach, P.~Benfield, T.~Helgaker and D.~J. Tozer, \emph{The Journal of
  chemical physics}, 2008, \textbf{128}, 044118\relax
\mciteBstWouldAddEndPuncttrue
\mciteSetBstMidEndSepPunct{\mcitedefaultmidpunct}
{\mcitedefaultendpunct}{\mcitedefaultseppunct}\relax
\EndOfBibitem
\bibitem[Pastore \emph{et~al.}(2012)Pastore, Mosconi, and
  De~Angelis]{Pastore2012}
M.~Pastore, E.~Mosconi and F.~De~Angelis, \emph{The Journal of Physical
  Chemistry C}, 2012, \textbf{116}, 5965--5973\relax
\mciteBstWouldAddEndPuncttrue
\mciteSetBstMidEndSepPunct{\mcitedefaultmidpunct}
{\mcitedefaultendpunct}{\mcitedefaultseppunct}\relax
\EndOfBibitem
\bibitem[Green \emph{et~al.}(2004)Green, Chandler, Haque, Nelson, and
  Durrant]{Green2004}
A.~N.~M. Green, R.~E. Chandler, S.~A. Haque, J.~Nelson and J.~R. Durrant,
  \emph{The Journal of Physical Chemistry B}, 2004, \textbf{109},
  142--150\relax
\mciteBstWouldAddEndPuncttrue
\mciteSetBstMidEndSepPunct{\mcitedefaultmidpunct}
{\mcitedefaultendpunct}{\mcitedefaultseppunct}\relax
\EndOfBibitem
\bibitem[Vittadini \emph{et~al.}(2000)Vittadini, Selloni, Rotzinger, and
  Grätzel]{Vittadini00}
A.~Vittadini, A.~Selloni, F.~P. Rotzinger and M.~Grätzel, \emph{The Journal of
  Physical Chemistry B}, 2000, \textbf{104}, 1300--1306\relax
\mciteBstWouldAddEndPuncttrue
\mciteSetBstMidEndSepPunct{\mcitedefaultmidpunct}
{\mcitedefaultendpunct}{\mcitedefaultseppunct}\relax
\EndOfBibitem
\bibitem[Nazeeruddin \emph{et~al.}(2003)Nazeeruddin, Humphry-Baker, Liska, and
  Grätzel]{Nazeeruddin2003}
M.~K. Nazeeruddin, R.~Humphry-Baker, P.~Liska and M.~Grätzel, \emph{The
  Journal of Physical Chemistry B}, 2003, \textbf{107}, 8981--8987\relax
\mciteBstWouldAddEndPuncttrue
\mciteSetBstMidEndSepPunct{\mcitedefaultmidpunct}
{\mcitedefaultendpunct}{\mcitedefaultseppunct}\relax
\EndOfBibitem
\bibitem[Srinivas \emph{et~al.}(2009)Srinivas, Yesudas, Bhanuprakash, ~, and
  Giribabu]{Srinivas09}
K.~Srinivas, K.~Yesudas, K.~Bhanuprakash, V.~J.~R. ~ and L.~Giribabu, \emph{The
  Journal of Physical Chemistry C}, 2009, \textbf{113}, 20117--20126\relax
\mciteBstWouldAddEndPuncttrue
\mciteSetBstMidEndSepPunct{\mcitedefaultmidpunct}
{\mcitedefaultendpunct}{\mcitedefaultseppunct}\relax
\EndOfBibitem
\bibitem[Hara \emph{et~al.}(2003)Hara, Sato, Katoh, Furube, Ohga, Shinpo, Suga,
  Sayama, Sugihara, and Arakawa]{Hara02}
K.~Hara, T.~Sato, R.~Katoh, A.~Furube, Y.~Ohga, A.~Shinpo, S.~Suga, K.~Sayama,
  H.~Sugihara and H.~Arakawa, \emph{The Journal of Physical Chemistry B}, 2003,
  \textbf{107}, 597--606\relax
\mciteBstWouldAddEndPuncttrue
\mciteSetBstMidEndSepPunct{\mcitedefaultmidpunct}
{\mcitedefaultendpunct}{\mcitedefaultseppunct}\relax
\EndOfBibitem
\bibitem[Politzer \emph{et~al.}(2007)Politzer, Lane, Concha, Ma, and
  Murray]{Politzer07}
P.~Politzer, P.~Lane, M.~Concha, Y.~Ma and J.~Murray, \emph{Journal of
  Molecular Modeling}, 2007, \textbf{13}, 305--311\relax
\mciteBstWouldAddEndPuncttrue
\mciteSetBstMidEndSepPunct{\mcitedefaultmidpunct}
{\mcitedefaultendpunct}{\mcitedefaultseppunct}\relax
\EndOfBibitem
\bibitem[Politzer \emph{et~al.}(2010)Politzer, Murray, and Clark]{Politzer10}
P.~Politzer, J.~S. Murray and T.~Clark, \emph{Phys. Chem. Chem. Phys.}, 2010,
  \textbf{12}, 7748--7757\relax
\mciteBstWouldAddEndPuncttrue
\mciteSetBstMidEndSepPunct{\mcitedefaultmidpunct}
{\mcitedefaultendpunct}{\mcitedefaultseppunct}\relax
\EndOfBibitem
\bibitem[Politzer \emph{et~al.}(2013)Politzer, Murray, and Clark]{Politzer13}
P.~Politzer, J.~S. Murray and T.~Clark, \emph{Phys. Chem. Chem. Phys.}, 2013,
  \textbf{15}, 11178--11189\relax
\mciteBstWouldAddEndPuncttrue
\mciteSetBstMidEndSepPunct{\mcitedefaultmidpunct}
{\mcitedefaultendpunct}{\mcitedefaultseppunct}\relax
\EndOfBibitem
\bibitem[Salvador \emph{et~al.}(2001)Salvador, Paizs, Duran, and
  Suhai]{Salvador01}
P.~Salvador, B.~Paizs, M.~Duran and S.~Suhai, \emph{Journal of Computational
  Chemistry}, 2001, \textbf{22}, 765--786\relax
\mciteBstWouldAddEndPuncttrue
\mciteSetBstMidEndSepPunct{\mcitedefaultmidpunct}
{\mcitedefaultendpunct}{\mcitedefaultseppunct}\relax
\EndOfBibitem
\end{mcitethebibliography}
\bibliographystyle{rsc} 
}

\end{document}